\documentclass[%
 aip,
 amsmath,amssymb,
preprint,%
]{revtex4-1}

\usepackage{graphicx}
\usepackage{dcolumn}
\usepackage{bm}

\usepackage[utf8]{inputenc}
\usepackage[T1]{fontenc}
\usepackage{mathptmx}

\usepackage{lineno}
\usepackage{float}
\usepackage{hyperref}
\usepackage{amsmath}
\usepackage{amssymb}
\usepackage{siunitx}
\providecommand{\e}[1]{\ensuremath{\times 10^{#1}}}
\newcommand*{\Rcavity}{\ensuremath{R_\text{cav}}}
\newcommand*{\Hcavity}{\ensuremath{L}}

\newcommand*{\TMone}{\ensuremath{\text{TM}_{010}\,}}
\DeclareSIUnit\year{yr}

\begin{document}

\preprint{AIP/123-QED}

\title[A Symmetric Multi-rod Tunable Microwave Cavity for the HAYSTAC Dark Matter Axion Search]{A Symmetric Multi-rod Tunable Microwave Cavity for the HAYSTAC Dark Matter Axion Search}

\author{Maria Simanovskaia}
 \email{simanovskaia@berkeley.edu.}
\author{Alex Droster}
\author{Heather Jackson}
\author{Isabella Urdinaran}
 \altaffiliation[Now at ]{Naval Air Station Pensacola.}
\author{Karl van Bibber}
\affiliation{ 
Department of Nuclear Engineering, University of California Berkeley, Berkeley CA, 94720 USA
}%

\date{\today}

\begin{abstract}
The microwave cavity experiment is the most sensitive way of looking for axions in the $0.1-\SI{10}{\giga \Hz}$ range, corresponding to masses of $0.5 - \SI{40}{\micro \eV}$. The particular challenge for frequencies greater than $\SI{5}{\giga \Hz}$ is designing a cavity with a large volume that contains a resonant mode that has a high form factor, a high quality factor, a wide dynamic range, and is free from intruder modes. For HAYSTAC, we have designed and constructed an optimized high frequency cavity with a tuning mechanism that preserves a high degree of rotational symmetry, critical to maximizing its figure of merit. This cavity covers an important frequency range according to recent theoretical estimates for the axion mass, $5.5 - \SI{7.4}{\giga \Hz}$, and the design appears extendable to higher frequencies as well. This paper will discuss key design and construction details of the cavity, present a summary of the design evolution, and alert practitioners of potentially unfruitful avenues for future work.
\end{abstract}

\maketitle

\section{Introduction}

The nature of dark matter is one of the most outstanding mysteries in science today. Possibilities for its composition range from black holes, to heavy particles that can be detected by their collisions with atomic nuclei, to light particles that act like a coherent field. Sensitive experiments have not yet been able to detect dark matter. A case can be made that the most promising dark matter candidate is the axion, which was invented to solve the strong charge-parity problem in the standard model of particle physics~\cite{Pec77a,Pec77b,Wei78,Wil78}. If the axion exists, it is likely to have been produced in abundance and constitute a major component of dark matter~\cite{Sik83b,Ste83}.

In the presence of a magnetic field, axions will be converted to an oscillating electromagnetic field with a frequency corresponding to the axion mass as $h\nu = mc^2$. The conversion is resonantly enhanced when the cavity mode of interest corresponds to this frequency and will manifest itself as a power excess inside the cavity~\cite{Sik83}. Many modes do not couple at all to the axion field; the one which provides the largest form factor is the \TMone mode. This must be a tuning search because the bandpass of the cavity is extremely narrow and the axion mass is unknown. Each microwave cavity can provide good sensitivity in the search for axions over a limited frequency range, but new technologies are necessary to probe the full axion parameter space in a reasonable amount of time~\cite{Gra15}. Both as a testbed for innovative cavity and amplifier concepts and as a data pathfinder for higher frequencies, the Haloscope At Yale Sensitive To Axion Cold dark matter (HAYSTAC) was the first cavity experiment to use a dilution refrigerator and a quantum-limited amplifier (a Josephson parametric amplifier), and most recently has operated with a squeezed-vacuum state receiver to circumvent the Standard Quantum Limit entirely~\cite{Bru16,Bru17,AlK17,Zho18}. 

The HAYSTAC Phase I and II cavity is a cylinder with an internal rod that rotates off-center and tunes the mode of interest's resonance frequency between 3.4 and $\SI{5.8}{\giga \Hz}$ in a volume of $\SI{1.5}{\liter}$. Several recent theoretical predictions for the axion mass motivate building cavities higher in frequency, up to $\SI{10}{\giga \Hz}$~\cite{KlaerMoore2017,Bus20}. The optimized cavity presented here will reach well-motivated parameter space.

\subsection{Expected signal power}

In a haloscope search such as HAYSTAC, the expected signal power is given by 
\begin{equation}\label{eq:power}
P_{\text{sig}} = \left(g_{a\gamma \gamma}^2 \frac{\hbar^3c^3\,\rho_a}{m_a^2}\right) \times \left(\frac{1}{\mu_0}B_0^2 \omega_c VC_{mn\ell}Q_L \frac{\beta}{1+\beta} \frac{1}{1+\left(2\Delta\nu_a/\Delta\nu_c\right)^2}\right),
\end{equation}
where the terms in the first set of parentheses involve unknown parameters of the particle physics of the axion and the dark matter, and the terms in the second set of parentheses are determined by experimental parameters. The unknown parameters (for which we have reasonable theoretical or observational estimates) include the model-dependent axion-photon coupling constant $g_{a\gamma \gamma} = \alpha g_\gamma\, m_a/  \pi \Lambda^2$, local dark matter density $\rho_a\approx \SI{0.45}{\giga \eV}/\text{cm}^3$ (commonly assumed in axion searches~\cite{Pen00} and consistent with recent measurements~\cite{Rea14}), and $\Lambda = \SI{78}{\mega \eV}$, which encodes the dependence of axion mass on hadronic physics. The relevant experimental parameters are external magnetic field strength $B_0$, cavity resonance frequency $\omega_c = 2 \pi \nu_c$, cavity volume $V$, mode-specific cavity form factor $C_{nml}$, cavity loaded quality factor $Q_L = Q_0/\left(1+\beta\right)$, cavity linewidth $\Delta \nu_c$, and $\beta$, which parameterizes the cavity coupling to the receiver ($\beta=1$ corresponds to critical coupling).

Typical values for the detector are $B_0 = \SI{9}{\tesla}$, $\omega_c = 2\pi \left( \SI{5}{\giga \Hz} \right)$, $V=\SI{1.5}{\liter}$, $C_{010} = 0.5$, $Q_L = 10^4$, $\beta = 2$, $\Delta \nu_a = \SI{5}{\kilo \Hz}$, and $\Delta \nu_c = \nu_c / Q_L = \omega_c / \left(2\pi Q_L\right)$. Altogether, the expected power for these parameters at $g_\gamma = -0.97$ (one representative value for the KSVZ family of models, frequently used as a benchmark value by experimentalists) is $P_\text{sig} \simeq 10^{-24}\,\text{W}$. 

\subsection{Noise considerations}

The Dicke radiometer equation \cite{Dic46} gives the signal-to-noise ratio $\Sigma$:
\begin{equation} \label{eq:SNR}
\Sigma = \frac{P_{\text{sig}}}{k_BT_{\text{sys}}}\sqrt{\frac{\tau}{\Delta\nu_a}},
\end{equation}
where $T_\text{sys}$ is the system noise temperature, $\tau$ is the integration time, and $\Delta \nu_a$ is the expected linewidth of the axion. It is much more effective to decrease noise $T_\text{sys}$ or increase power $P_\text{sig}$ rather than to improve $\Sigma$ by longer integration times.

For any phase-insensitive linear receiver the system noise temperature $T_{\text{sys}}$ may be written
\begin{equation}\label{eq:noise}
k_BT_{\text{sys}} = h\nu N_{\text{sys}} = h\nu\left(\frac{1}{e^{h\nu/k_BT} -1} + \frac{1}{2} + N_A\right),
\end{equation}
where the three additive contributions correspond respectively to a blackbody photon spectrum characteristic of the cavity at temperature $T$, the zero-point fluctuations of the cavity, and the input-referred added noise of the receiver. The latter two terms combine to form the Standard Quantum Limit (SQL), i.e. the irreducible noise associated with a linear amplifier, $h\nu = k_BT$. During Phase I, HAYSTAC used one quantum-limited JPA as an amplifier. In the ongoing Phase II operation, a squeezed state receiver with two JPAs has been incorporated, which can circumvent the SQL entirely.   Here the first JPA prepares the cavity in a squeezed vacuum state before amplification; the second JPA amplifies the output from the cavity, with the putative axion signal superposed on the squeezed vacuum state. HAYSTAC has demonstrated that the squeezed state receiver improves the signal-to-noise relative to a single JPA linear amplifier~\cite{Mal19}.

Increasing power by increasing the applied magnetic field or improving cavity parameters would also improve the signal-to-noise ratio. 

\subsection{Scan rate}

The ultimate figure of merit of HAYSTAC is the scan rate, which incorporates the expected signal power and noise considerations, and quantifies the scan speed through different frequencies at a given sensitivity

\begin{equation}\label{eq:scan}
R \equiv \frac{\mathrm{d}\nu}{\mathrm{d}t} \approx \frac{4}{5}\frac{Q_LQ_a}{\Sigma^2} \left(g_{a\gamma \gamma}^2\frac{\hbar^3c^3\rho_a}{m_a^2} \right)^2 \times \left(\frac{1}{\hbar\mu_0}\frac{\beta}{1+\beta}B_0^2VC_{mn\ell}\frac{1}{N_{\text{sys}}}\right)^2.
\end{equation}

Most of these terms appear in the expression for signal power in Eq.~\ref{eq:power}, and $\Sigma$ is the signal-to-noise ratio in Eq.~\ref{eq:SNR}. If our goal in sensitivity was modest, for example $2g^\text{KSVZ}_{a\gamma \gamma}$, in its Phase I configuration HAYSTAC would achieve a scan rate of $\SI{600}{\mega \Hz / \year}$. However, as the scan rate is proportional to the fourth power of the coupling constant, to achieve a sensitivity of $g^\text{KSVZ}_{a\gamma \gamma}$, the scan rate reduces to a mere $\SI{40}{\mega \Hz / \year}$, for which a feasible search is questionable.      

We can improve the scan rate by buying or building a magnet with a stronger magnetic field and larger volume, decreasing noise in the receiver, or improving cavity parameters. These improvements allow us to search through a mass range more quickly, but we are still limited by the tuning range of the cavity and amplifier electronics. The frequency range we can probe depends on the resonance frequency of the microwave cavity. Specifically, we are interested in accessing higher frequencies since recent theoretical work favor higher-mass axions as mentioned previously. In general, higher frequency cavities have a smaller volume and therefore suffer from a smaller expected signal power as well as an increase in operational complexity due to a higher intruder mode density. New cavity designs can expand the accessible frequency range while improving sensitivity. This requires investigation of various geometries using electromagnetic simulations, prototypes, and microwave testing. 

\section{Required Qualities of the Cavity} 

The cavity figure of merit is defined by maximizing the scan rate at a given sensitivity and is determined by a combination of physical characteristics of the cavity (geometry, conductivity, etc.) for a resonant mode chosen for operation. The components include the quality factor $Q$, which quantifies losses, the form factor $C$, which is determined by the overlap of the resonant mode's electric field with the external magnetic field, and the cavity volume $V$. The scan rate depends on these quantities as
\begin{equation} 
R \equiv \frac{d \nu}{dt} \propto Q\, C^2\, V^2.
\label{eq:cavityFOM}
\end{equation}
HAYSTAC aims to optimize cavities for this figure of merit while considering mode purity throughout the tuning range and the frequency range of interest. Each cavity geometry can support an infinite number of modes of various electric and magnetic field profiles. Some of these modes are transverse electric (TE) modes, which interfere with the transverse magnetic (TM) mode of interest intermittently throughout its tuning range. The cavity figure of merit applies to a single resonant mode with its associated resonance frequency, which is related to the axion mass by $E = m_a c^2 = h \nu$. The axion mass $m_a$ corresponding to a resonance frequency $\nu$ is approximately
\begin{equation}
m_a \left[\SI{}{\micro\eV}\right] = 4.136\, \nu \left[\text{GHz}\right].
\end{equation}
Changing the cavity geometry (for example, by moving a tuning rod) changes the mode frequencies. Mode maps track these changes by showing mode frequencies at each cavity geometry. 

\subsection{Quality factor}
The quality factor is determined primarily by the material and geometry of the cavity, and it also depends on the resonant mode. Materials with a higher conductivity have smaller skin depths and therefore higher quality factors. Plating the cavities with oxygen-free high-conductivity (OFHC) copper and annealing them increases conductivity.

The quality factor of a cavity is given by the ratio of the stored energy $U$ to the dissipated power $P_d$, multiplied by the resonant mode frequency $\omega$: 
\begin{equation}
Q = \omega \, \frac{U}{P_d}.
\label{eq:qualityfactor}
\end{equation}

The stored energy in the cavity is proportional to the square of the electric field integrated over the cavity volume
\begin{equation}
U = \frac{1}{2} \epsilon_0 \int_\text{cavity\,volume}{|\vec{E}|^2}\,dV.\\
\end{equation}

The power loss in the cavity is proportional to the square of the magnetic field integrated over the metallic surfaces inside the cavity. Then,
\begin{equation}
P_d =\frac{ \omega\, \mu \, \delta}{4} \int_\text{cavity\,surfaces}{ |\vec{H}|^2}\,dA, \\
\end{equation}
where the skin depth $\delta$ is the distance that electric fields are allowed to penetrate into the metallic surfaces. The classical skin depth is given by
\begin{equation}
\delta = \sqrt{\frac{2}{\omega\, \mu \, \sigma}}.
\label{eq:skindepth}
\end{equation}
Since the conductivity $\sigma$ improves with decreasing temperature, the classical skin depth is expected to improve as well. However, at sufficiently low temperatures, the skin depth reaches an asymptote. In HAYSTAC, cooling the cavities from room temperature to $\SI{4}{\kelvin}$ improves the quality factor by approximately a factor of four at a frequency around $\SI{1}{\giga \Hz}$. For comparison, the conductivity $\sigma$ improves by a factor of over a hundred in that temperature range. The classical behavior becomes invalid when the skin depth decreases below the electron's mean free path. In this regime, the skin depth depends on the electron density instead of the normal conductivity. This anomalous skin depth~\cite{Pip47} is given by 
\begin{equation}
\label{eq:anomskindepth}
\delta_a = \left( \frac{\sqrt{3}\, c^2 m_e v_F}{8\pi^2 \omega n e^2} \right)^{1/3},
\end{equation}
where $m_e$ is the electron mass, $v_F$ is the Fermi velocity, $n$ is the conduction electron density, and $e$ is the electron charge~\cite{Kit87}.

\subsection{Form factor}
The form factor is determined by the overlap of the electric field of the mode $\vec{E}_{nml}=\vec{E}$ with the external magnetic field $\vec{B}_0= B_0 \hat{z}$. It is given by the equation
\begin{equation}
C_{nml}  = \frac{ \left(\int{\vec{E} \cdot \vec{B}_0}\,dV \right)^2}{B_0^2 \, V \int{\epsilon_r E^2}\,dV} =  \frac{\left(\int{E_z B_0}\,dV \right)^2}{B_0^2 \, V \int{\epsilon_r E^2}\,dV}  = \frac{\left(\int{E_z}\,dV \right)^2}{V \int{\epsilon_r E^2}\,dV},
\label{eq:formfactor}
\end{equation}
where $L$ is the cavity length, and $V$ is the cavity volume not occupied by a metallic object.

The form factor $C_{nml}$ is maximized when the integral of $\vec{E}_{nml} \cdot \vec{B}_0$ is maximized. Since the applied magnetic field in HAYSTAC is in the $\hat{z}$ direction, all TE modes, all transverse electromagnetic (TEM) modes, and TM$_{nml}$ with either $n>0$ or $l>0$ have form factors that are identically zero.

\subsection{Volume}

The volume in the cavity figure of merit involves the internal cavity space through which electric fields penetrate. It is the same volume $V$ as in the form factor calculation in Eq.~\ref{eq:formfactor}. For the HAYSTAC sensitivity calculations, the volume includes the space involving vacuum and dielectric materials, but excludes metal pieces \footnote{Although this volume definition differs from a previously-used definition, the scan rate calculations are equivalent since volume only appears as a product with the form factor. Other experiments defined volume as the space inside the cavity outer metal boundary, including the metal pieces as well as the space involving vacuum and dielectric materials.}. Preserving the mode, cavity geometry, and cavity aspect ratio leads to a volume depending on frequency as $\nu^{-3}$. This volume decrease with increasing frequencies is one of the main challenges in designing higher-frequency cavities. 

\subsection{Mode density}

\begin{figure}
\includegraphics{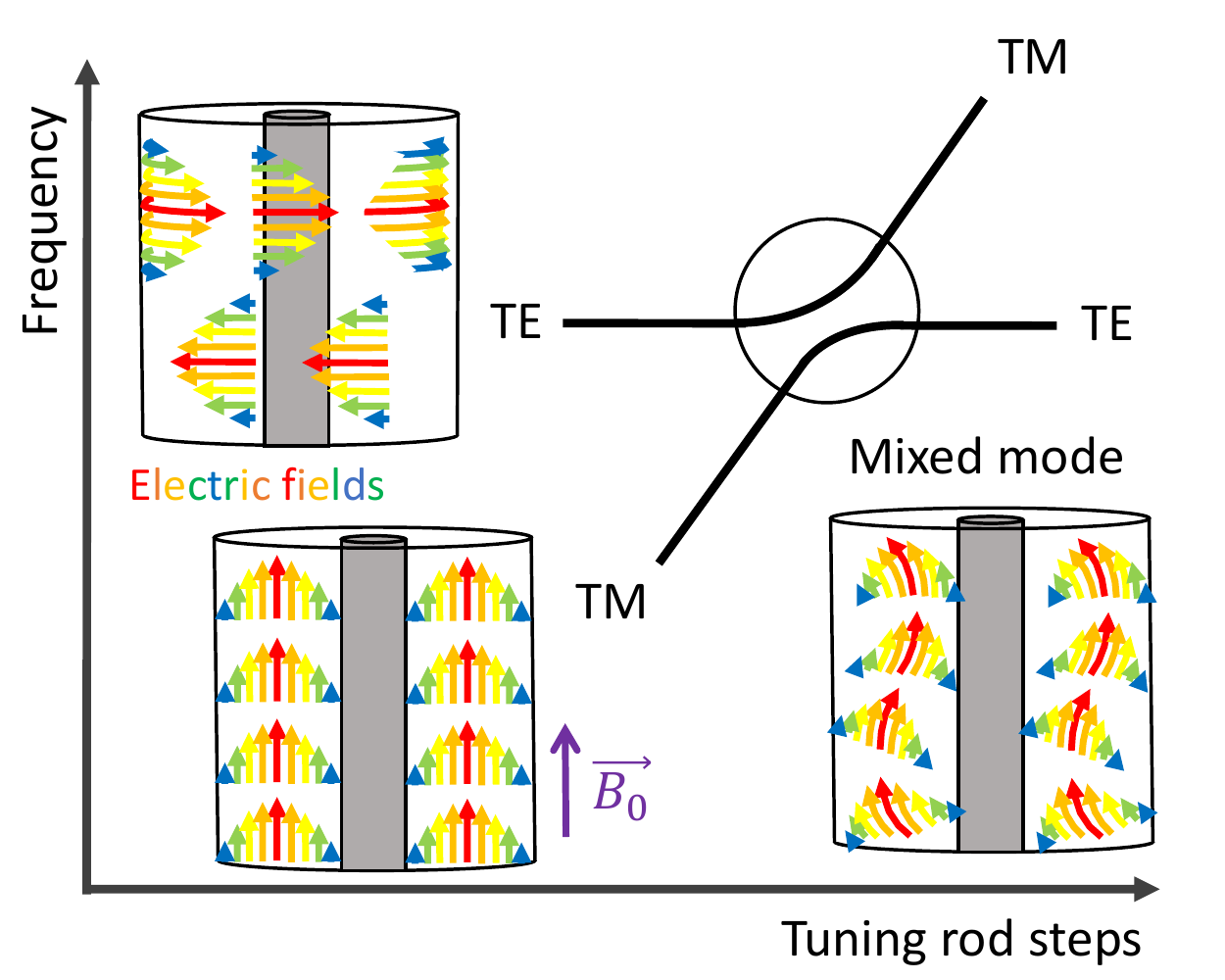}
\caption{\label{fig:modemixing} Electric field profile cartoons of a TE mode, TM mode, and a mixed mode in a microwave cavity; the grey cylinder represents a tuning rod. The small violet arrow represents $\vec{B_0}$, the external applied magnetic field. The plot shows the effect of rod rotation on resonance mode frequency. When the TE and TM modes are sufficiently close in frequency, they mix and create a mixed hybrid mode.} 
\end{figure}

Quality factor, form factor, and volume quantitatively describe the behavior of interest of a single resonant mode at a given frequency. These quantities are useful for quantifying general performance across a tuning range, but if the range is full of intruder modes, there will be gaps in the frequency coverage of the search, which will need to be filled in afterwards by one means or another, with the result that the actual scan rate will be much slower than predicted by Eq.~\ref{eq:scan}. Figure~\ref{fig:modemixing} shows the electric field profiles of three example resonant modes labeled ``TE'', ``TM'', and ``mixed mode''. The TM mode resonance frequency is denoted by the black curve, and the frequency increases with increasing tuning rod steps. In comparison, the TE mode resonance frequency does not change significantly. When the TE and TM mode frequencies approach each other, the two modes mix, producing two hybrid modes, in complete analogy with two-level mixing in quantum mechanics. If the mode of interest hybridizes significantly, it will be difficult or impossible to interpret the results of the experiment, thus leading to a notch in frequency coverage of the experiment. Mode density is difficult to quantify, but it is a key consideration for cavity design. The intruder mode density problem worsens for cavities of too large an aspect ratio $\Hcavity/\Rcavity$; practically one is constrained to stay with cavity designs of $\Hcavity/\Rcavity \sim 5$ or lower.

\section{Cavity designs}
\label{sec:cav_designs}

The cavity barrel size is constrained in width by the size of the magnet bore and in length by the density of TE and TEM modes in our frequency range of interest. To probe higher frequencies with a \TMone-like mode, the rod size must increase. Increasing the rod size exaggerates the electric field asymmetry when the rod is off-center and decreases the cavity volume. 

\subsection{One-rod cavity}
\begin{figure}
	\includegraphics{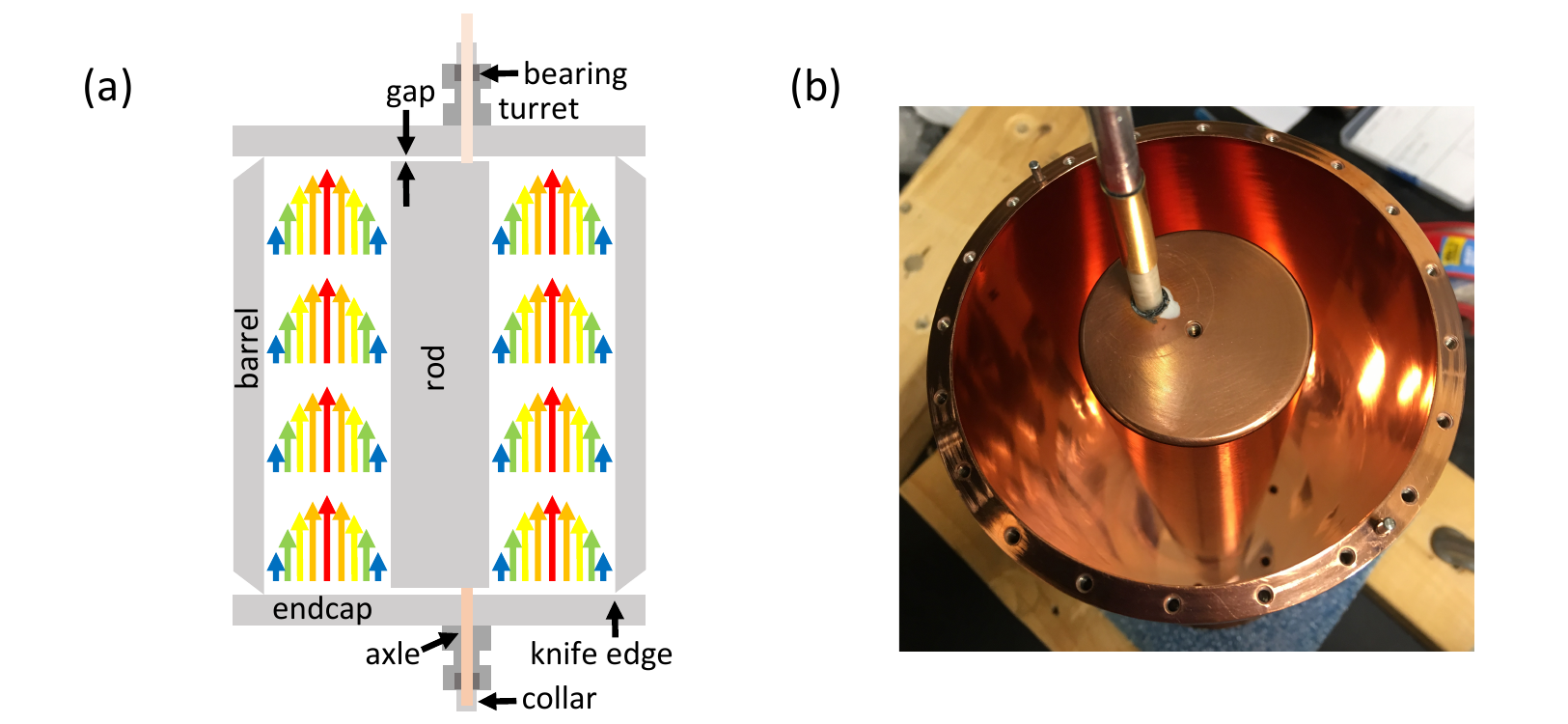}
	\caption{\label{fig:1rod} One-rod cavity (a) cartoon image with spatially-dependent electric field strength of the \TMone-like mode (electric field intensity ranges between maxima denoted by long red arrows and minima denoted by short blue arrows) and (b) top-view photographic image with top endcap removed.} 
\end{figure}

HAYSTAC Phases I and II used the \TMone-like mode of a copper-coated stainless steel one-rod cavity, as shown by the simplified schematic and photograph in Fig.~\ref{fig:1rod}. The barrel has inner radius $2''$ and height $10''$, and the rod has radius $1''$ and height $9.98''$, making the gaps between the rod and endcaps $0.01''$ on each side. The rod rotates around an axle which is partially composed of $0.25''$ outer-diameter (OD) alumina tubes centered $0.475''$ away from the cavity and rod centers. The axles extend through the endcaps, turrets, bearings, and collars. The bearings provide nearly frictionless rod rotation and the collars fix the gaps on either side of the rod, otherwise the rod would rest on the bottom endcap. We use two dowel pins on either side of the barrel to align the endcaps. This cavity was experimentally characterized in detail~\cite{Rap19}.

\begin{figure}
	\includegraphics[width=\textwidth]{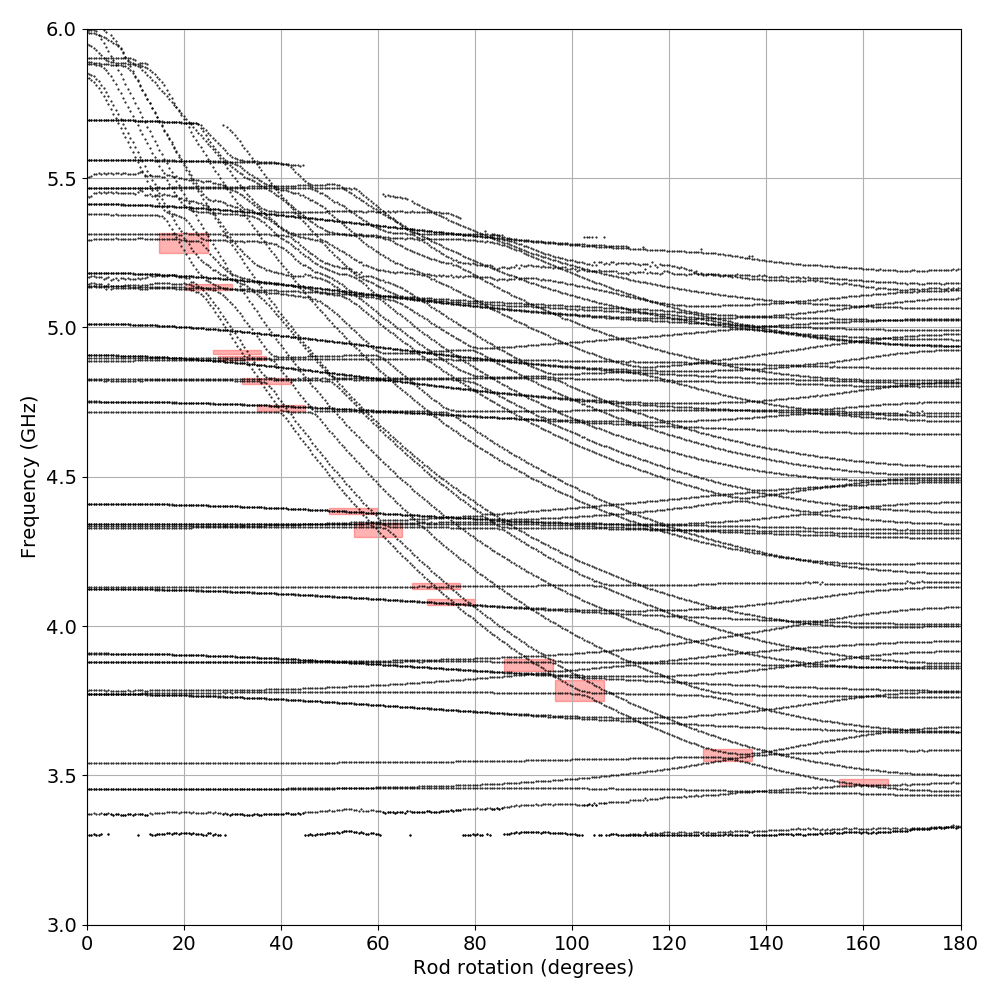} 
	\caption{Simulated mode map shows a black dot at each possible mode frequency at discrete rod locations in a cavity matching the one used in HAYSTAC Phases I and II. TM mode frequencies change significantly with rod rotation compared to those of TE and TEM modes. \TMone-like mode frequencies correspond to the lowest-frequency TM mode at all rod rotations. Red boxes correspond to experimentally-observed mode crossings of the \TMone-like mode. The box heights reflect mode crossing ranges observed during cavity characterization, while box lengths and horizontal positions are arbitrary.}
	\label{fig:1rod_modemap}
\end{figure}

Simulations of the Phase I and II one-rod HAYSTAC cavity can predict the resonance frequencies of all possible modes and mode crossings, as in the simulated mode map in Fig.~\ref{fig:1rod_modemap}. As the rod is rotated away from the center of the cavity in steps of half a degree, the resonance mode frequencies increase, decrease, or stay constant depending on the mode profile. Throughout the full rotation, the \TMone-like mode tunes in the range $3.4-\SI{5.8}{\giga \Hz}$, mixing with a few modes in the frequency ranges highlighted by red boxes in Fig.~\ref{fig:1rod_modemap}. In this cavity, TE mode crossings were found to be much wider than TEM mode crossings~\cite{Rap19}. As mentioned previously, notches due to mode crossings ultimately must be filled in to attain complete coverage of the mass range. One way of accomplishing this is by insertion of a small dielectric rod longitudinally in the cavity; this serves as a vernier control allowing us to shift both the TE and TM modes slightly downward in frequency, thus moving the avoided crossing away from the region of interest, so it can now be filled in.

Computer Simulation Technology Microwave Studio (CST MWS)~\footnote{Computer Simulation Technology Microwave Studio, https://www.cst.com/products/cstmws} simulations involve a copper rod with $0.010''$ gaps on either side with $0.25''$ OD alumina axles at the pivot point surrounded by vacuum and then a copper background marking the inner surface of the cavity. Note that every predicted mode crossing in the mode map corresponds to a decrease in \TMone-like mode form factor. Some mode crossings are too narrow to cause significant disruption in the tuning range, and we do not see them appear in the experimentally observed mode crossing red regions. It is practically impossible to experimentally measure the form factor of a resonant mode, so simulated form factors enter into our calculated sensitivity. Simulations can carefully map the frequency-dependent form factor.

\subsection{Two-rod cavity designs}

Simulations of the \TMone-like mode in the one-rod cavity indicate that the form factor is maximized in aximuthally-symmetric geometries, i.e. when the rod is in the center of the cavity. Since the quality factor behavior is difficult to predict, we can motivate our designs by improving the form factor and volume. Based on this concept, we would optimally have a rod in the center that changes in radius without breaking symmetry. The itinerary of our multi-rod design study throughout this project was guided by the principle that the best concept would likely approximate this ideal. Indeed the cavity that was ultimately constructed possessed discrete six-fold azimuthal symmetry, but we feel it may be instructive to show some other designs that initially looked promising, but proved unsatisfactory. 

\begin{figure}
	\includegraphics[width=.75\textwidth]{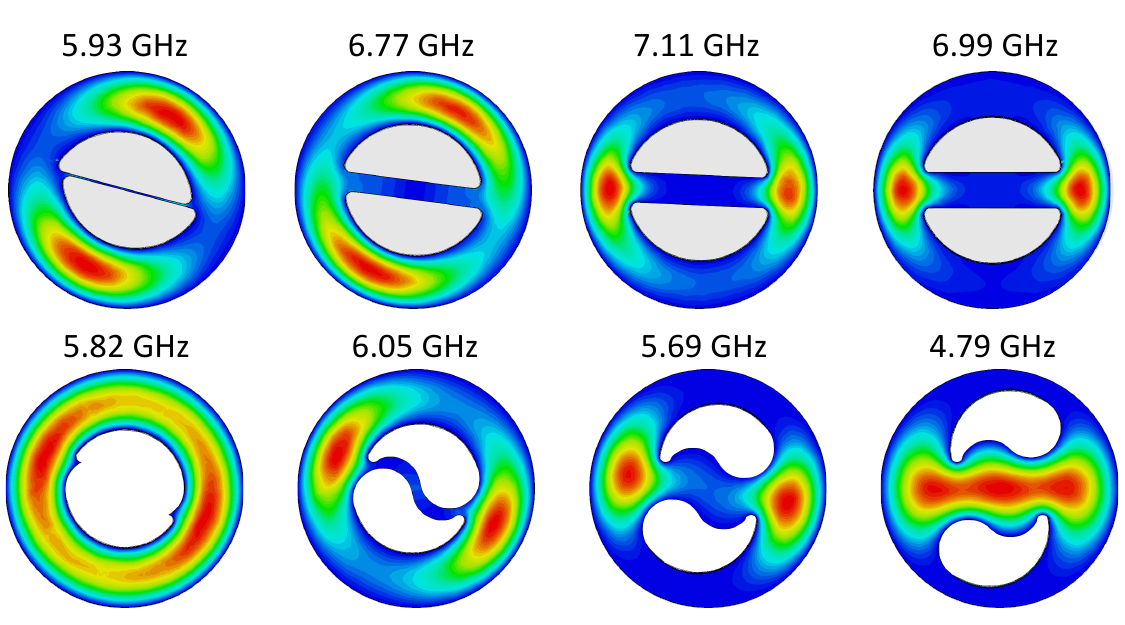}
	\caption{Electric field intensity plots for different rod rotations of two-rod cavity designs, with the \TMone-like mode resonance frequency noted above each corresponding image.}
	\label{fig:2rodphotos}
\end{figure}

One concept involved dividing the rod in two pieces and tuning the \TMone-like mode resonance frequency by pivoting the two rods away from each other. The two designs we considered were colloquially called the ``yin-yang'' and ``biscotti'' cavities. We simulated both designs with $0.010''$ gaps and without axles to find the cavity figure of merit across the tuning ranges. We originally expected the \TMone-like mode frequency to increase as the rods moved apart and as the space between the rod and cavity wall shrunk, the two rods together maintaining a roughly circular cross section. However, we observed that the \TMone-like mode resonance frequency decreases as the electric field begins to occupy the space between the rods instead. The heat-map plots of electric field in Fig.~\ref{fig:2rodphotos} demonstrate this behavior for both configurations as a function of the split-rod separation. Although the performance of these two cavities is approximately the same as that of the familiar one-rod cavity, these two-rod designs do not provide a promising concept for accessing higher frequencies.

\subsection{Seven-rod cavity}

\begin{figure}
	\includegraphics{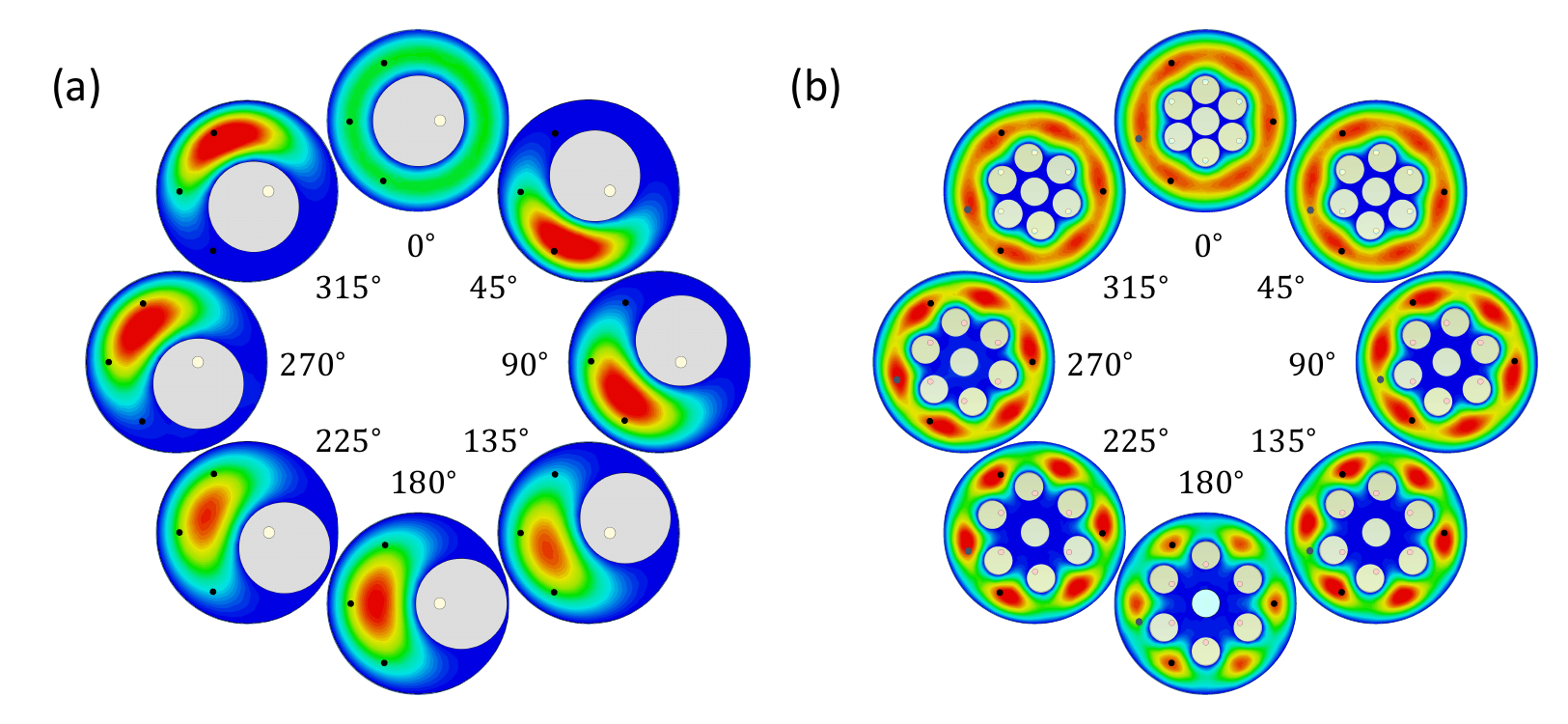}
	\caption{\label{fig:7rodrotations} Top-view heat maps of \TMone-like mode electric field strength at various rod rotations in the (a) one $2''$ OD rod and (b) seven $0.625''$ OD rod design. Three holes available for antennas are marked by black dots.} 
\end{figure}

In a seven-rod design, the six peripheral rods pivot in unison symmetrically away from the central rod. The central rod keeps the electric field of the $\TMone$-like mode from entering the space between the rods. Figure~\ref{fig:7rodrotations} shows heat maps of the \TMone-like mode electric field magnitude for several rod rotations in the seven-rod design. For comparison, the one-rod design heat maps are in Fig.~\ref{fig:7rodrotations}a. The simulated heat maps shown in Fig.~\ref{fig:7rodrotations} are cross sections taken at the center of the $4''$ diameter cavity and include seven $0.625''$ OD rods. The $0^\circ$ rotation in the seven-rod design, where the seven rods are closest to each other, corresponds to the minimum frequency of the \TMone-like mode, while the $0^\circ$ rotation in the one-rod design, where the rod is in the center of the cavity, corresponds to the maximum frequency of the \TMone-like mode in that design. These rod rotations give the most rotationally-symmetric electric field shape. Throughout the whole tuning range, the seven-rod design preserves perfect six-fold symmetry, whereas the one-rod design breaks all rotational symmetry at all angles once it has pivoted away from the origin.

\begin{figure}
	\includegraphics{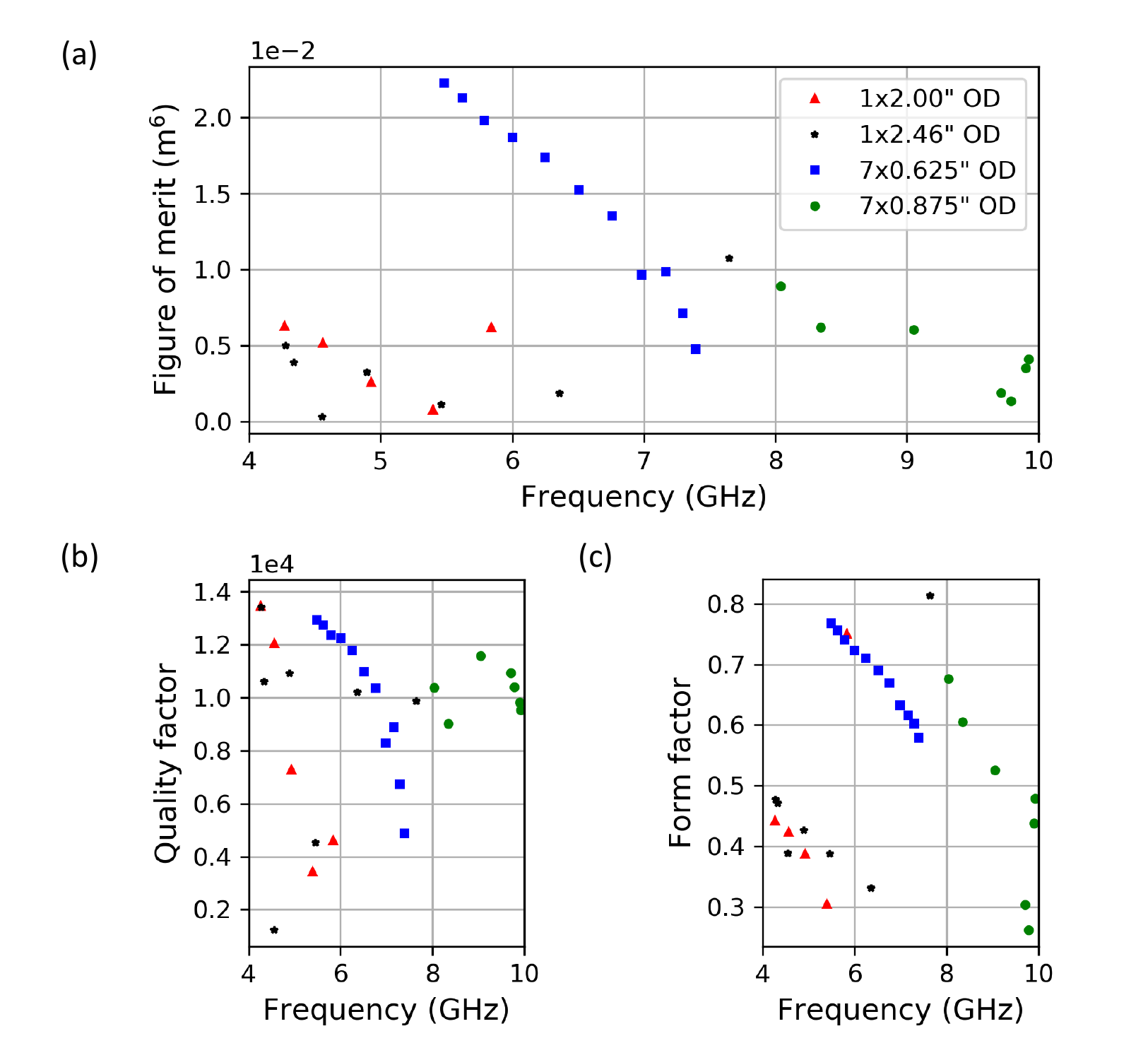}
	\caption{\label{fig:7rodfom} Comparison of \TMone-like mode behavior in seven-rod and one-rod cavity designs. Plots of (a) figure of merit $QC^2V^2$, (b) quality factor $Q$, and (c) form factor $C$. Simulations take into account $0.01''$ gaps on either side of the rods and appropriately-sized alumina axles.}
\end{figure}

Figure~\ref{fig:7rodfom} and Table~\ref{tbl:comparison} compare the simulated performance of one-rod cavities to that of seven-rod cavities, both incorporating $0.01''$ gaps between the rods and endcaps as well as appropriately-sized alumina axles. In case of later comparisons with measured data, the final gap sizes in the seven-rod cavity are much smaller than $0.01''$, which in some cases correspond to better performance than the simulations with larger gaps. To compare cavity designs, we focus on the figure of merit $QC^2V^2$, its components, and the accessible frequency range of the \TMone-like mode in the designs. As presented in Table~\ref{tbl:comparison}, the \TMone-like mode in a seven-rod design reaches a higher frequency in a larger volume compared to the one-rod design. This behavior of larger volumes with the seven-rod design could be extended higher in frequency.

\begin{table}
\caption{Comparison of cavity volume and maximum frequency of the \TMone-like mode for one-rod and seven-rod designs.}
\centering
\begin{tabular}{|c|c|c|}\hline 
	Design & Volume & Maximum frequency \\ \hline
	$1\times 2''$ OD & $1.55\e{-3}\text{ m}^3$ &  $\SI{5.8}{\giga \Hz}$ \\
	$7\times 0.625''$ OD & $1.71\e{-3}\text{ m}^3$ &  $\SI{7.4}{\giga \Hz}$ \\
	$1\times 2.46''$ OD & $1.37\e{-3}\text{ m}^3$ &  $\SI{7.8}{\giga \Hz}$ \\
	$7\times 0.875''$ OD & $1.28\e{-3}\text{ m}^3$ &  $\SI{9.9}{\giga \Hz}$ \\ \hline
\end{tabular}
\label{tbl:comparison}
\end{table}%

The plots in Fig.~\ref{fig:7rodfom} show the figure of merit $QC^2V^2$ variation across the \TMone-like mode frequency range in one $2''$ OD rod, one $2.46''$ OD rod, seven $0.625''$ OD rods, and seven $0.875''$ OD rods cavity designs. The form factor is maximized in the $0^\circ$ rotation for all designs, which correspond to the top of the frequency range in the one-rod designs and to the bottom of the frequency range in the seven-rod designs, respectively. As suggested by the symmetry in the seven-rod design throughout the tuning range, the form factor only slightly decreases as the \TMone-like mode is confined to smaller volumes at higher frequencies. In contrast, the form factor for most of the tuning range in a one-rod design is significantly lower than its maximum. Overall, other than the instances when the single rod is in the middle of the cavity, the seven-rod designs give a higher figure of merit than the one-rod designs up to at least around $\SI{9}{\giga \Hz}$. Higher frequencies are difficult to obtain since in the seven rod case, the electric field begins to occupy the space between the central rod and six moving rods. To avoid this and to keep increasing \TMone-like mode frequencies, we must turn to larger rods, which limit our tuning range.

\subsection{Plating and baking}
To improve electromagnetic performance and to decrease the gaps between the rods and endcaps, we copper-plated the seven-rod cavity with a thickness of several skin depths. The skin-depth of copper at $\SI{5}{\giga \Hz}$, given by Eq.~\ref{eq:skindepth} and using standard parameters, is approximately $\delta \sim \SI{1}{\micro \meter}$. The anomalous skin-depth of copper at $\SI{5}{\giga \Hz}$, given by Eq.~\ref{eq:anomskindepth}, is approximately $\delta_a \sim \SI{5.6}{\micro \meter}$. Therefore, a copper layer of thickness $0.006''$ (around $\SI{150}{\micro \meter}$) is sufficient.

After plating, baking the copper should improve its electromagnetic performance by lowering its resistivity. For this purpose, the rods and barrel were heated to $270^\circ \text{C}$ and the endcaps were heated to $350^\circ \text{C}$ in an $H_2$ environment. Although the cavity was not tested at each stage of manufacturing, the quality factor of the empty cavity after plating and annealing was approximately $2.74\e{4}$ at weak coupling, compared to the theoretically-predicted $3.07\e{4}$.

\subsection{Mechanical Design}

\begin{figure} 
	\includegraphics{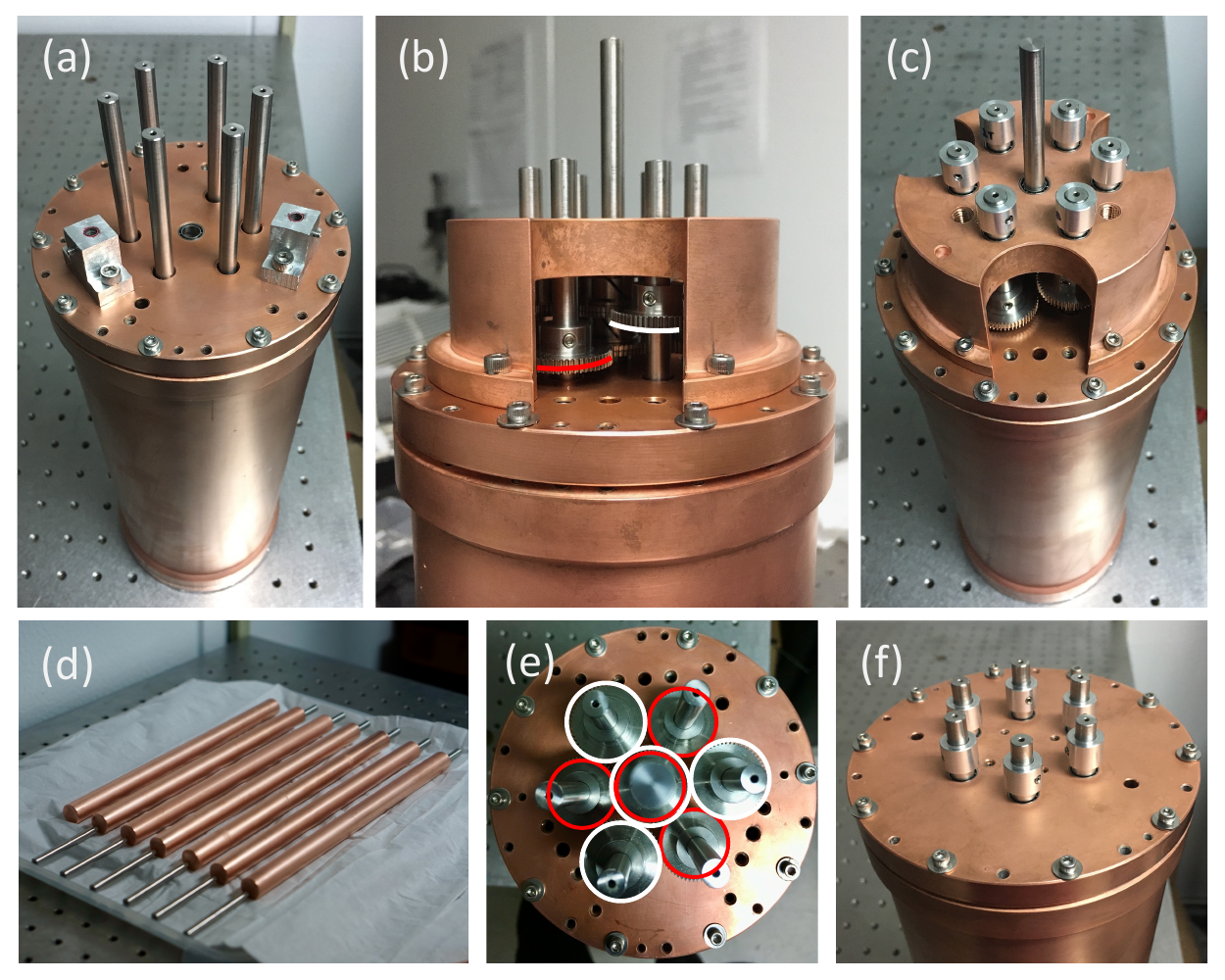} 
	\caption{\label{fig:7rodphotos} Seven-rod cavity in various stages of assembly, including (a) top-view with antenna ports, (b) top-view with gears and gear case, (c) top-view with gears, gear case, and collars, (d) rods outside of the cavity, (e) top-view with gears, and (f) bottom-view with collars. The red and white colors in (b) and (e) outline the two sets of gears, each containing three peripheral gears connected to an anti-backlash gear on the central rod.}	
\end{figure}

The six peripheral rods are connected via gears to a central rod. Rotating the central rod rotates the six peripheral rods symmetrically. Figure~\ref{fig:7rodphotos} shows the different components of the seven-rod cavity after copper-plating and re-machining. The peripheral rods connect to alumina axles on either side, which in turn connect to stainless steel rods. The six peripheral rods and the central rod are shown in Fig.~\ref{fig:7rodphotos}d. The stainless steel tops of these rods appear in Fig.~\ref{fig:7rodphotos}a, and hold the gears in Figs.~\ref{fig:7rodphotos}b, \ref{fig:7rodphotos}c, and \ref{fig:7rodphotos}e. The stainless steel bottoms of the peripheral rods appear with collars in Fig.~\ref{fig:7rodphotos}f.

The peripheral gears are stainless steel spur gears with $56$ teeth and $0.875''$ pitch diameter. These peripheral gears mesh in two sets of three, each to an anti-backlash gear on the central rod. All gears were purchased from Stock Drive Products / Sterling Instrument~\footnote{Stock Drive Products / Sterling Instrument, https://sdp-si.com/}. Figures ~\ref{fig:7rodphotos}b and \ref{fig:7rodphotos}e highlight the two sets of three peripheral gears, one set outlined in white along with the anti-backlash gear on the central rod and the other set outlined in red. The two sets exist at different heights on the central rod to prevent collisions. The central rod pivots all peripheral rods at the same time. 

\subsection{Cavity assembly}
Assembly involves a torque wrench for attaching the endcaps to the barrel and tightening the screws in a star pattern to avoid damaging the fine threads and to ensure a good seal with the barrel knife edge. Another key aspect of assembling the cavity is taking steps to align the rod correctly. This requires using an alignment tool to keep rods in the correct rotation and fixing the rod axles in the bearings before engaging the gears. The final coated and re-machined parts leave approximately $0.007 \pm 0.001''$ total for the two gaps between the rods and endcaps, which was measured using leaf gauges. Overall, the rotation is smooth throughout the tuning range except a slight resistance in parts most likely due to the tight fit and slight deformation of the bearings.

\section{Testing and characterization}
\label{sec:7rod_testing}

\begin{figure}
	\includegraphics{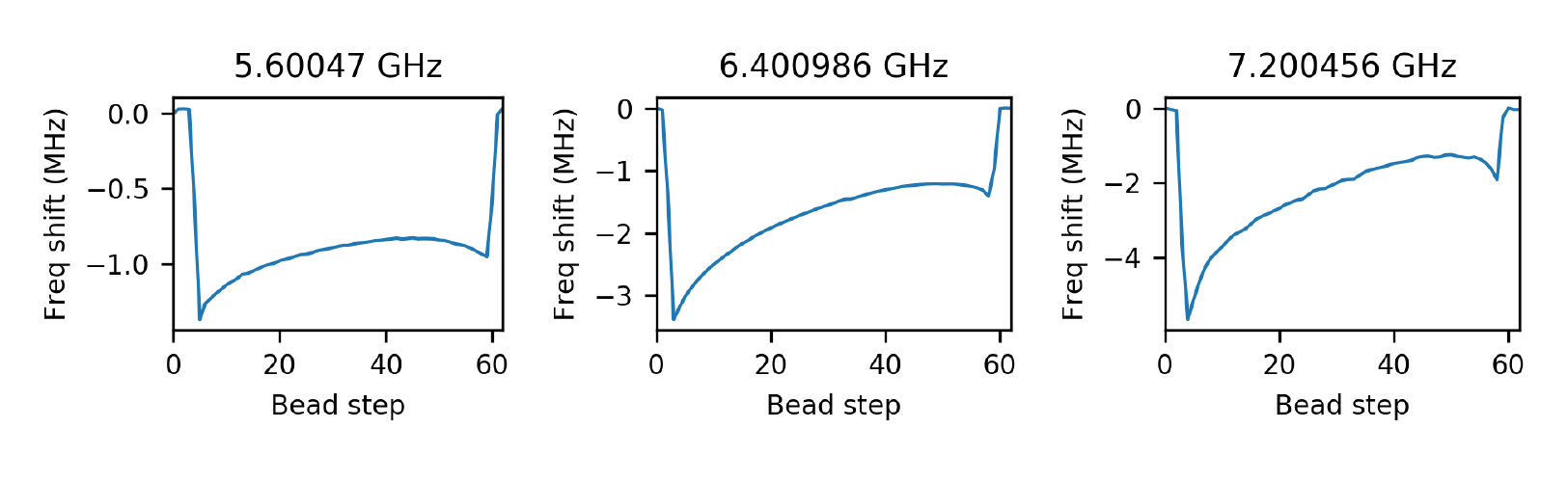}
	\caption{\label{fig:BP2} Measured frequency shift of the \TMone-like mode from bead perturbations at steps along the length of the seven-rod cavity. The plot title specifies the unperturbed \TMone-like mode resonance frequency corresponding to a frequency shift of zero. The dielectric bead decreases the resonance frequency when it enters the cavity.}\qquad 
\end{figure}

We can observe the frequencies and quality factors of resonant modes in our cavity by measuring the $S_{21}$ scattering parameter, which corresponds to the transmission from port 1 to port 2. All measurements are done with weak coupling to our coaxial antennas to minimize perturbations due to the antenna presence in the cavity. In our existing set-up, we can calibrate effects of cables, but unfortunately not the effects of reflections from within the antennas. Scattering parameter measurements give us the frequencies and quality factors of resonant modes, but not the distribution of electric field within the cavity. The electric field distribution helps characterize the resonant mode purity, which affects our sensitivity. To observe the resonant mode shape, we pull a small bead through the length of the cavity, measuring the resonance frequency at each step. The presence of the bead inside the cavity perturbs the electromagnetic field and shifts the resonance frequency by a magnitude proportional to the square of the strength of the electric field at the bead location~\cite{AlK17,Mai52}. If the bead is only slightly perturbing the electric field, the frequency shift we expect is given by
\begin{equation}
\frac{\Delta \omega}{\omega} = \frac{-\left(\epsilon -1 \right)}{2} \frac{V_\text{bead}}{V_\text{cav}} \frac{E(r)^2}{\left< E(r)^2 \right>_\text{cav}},
\end{equation}
where $V_\text{bead}$ and $V_\text{cav}$ are the volumes of the bead and cavity, respectively~\cite{Sla46}. This bead perturbation technique is commonly used in the microwave engineering community, but HAYSTAC has introduced its use in the microwave cavity axion search~\cite{AlK17,Rap19}. Bead perturbation measurement results at several rod locations of the seven-rod cavity are shown in Fig.~\ref{fig:BP2} for rods rotating in one direction. The larger frequency shifts at the edges of the cavity are evidence of mode localization most likely due to the gaps between the rods and the endcaps~\cite{Rap19}.

\begin{figure} 
	\includegraphics{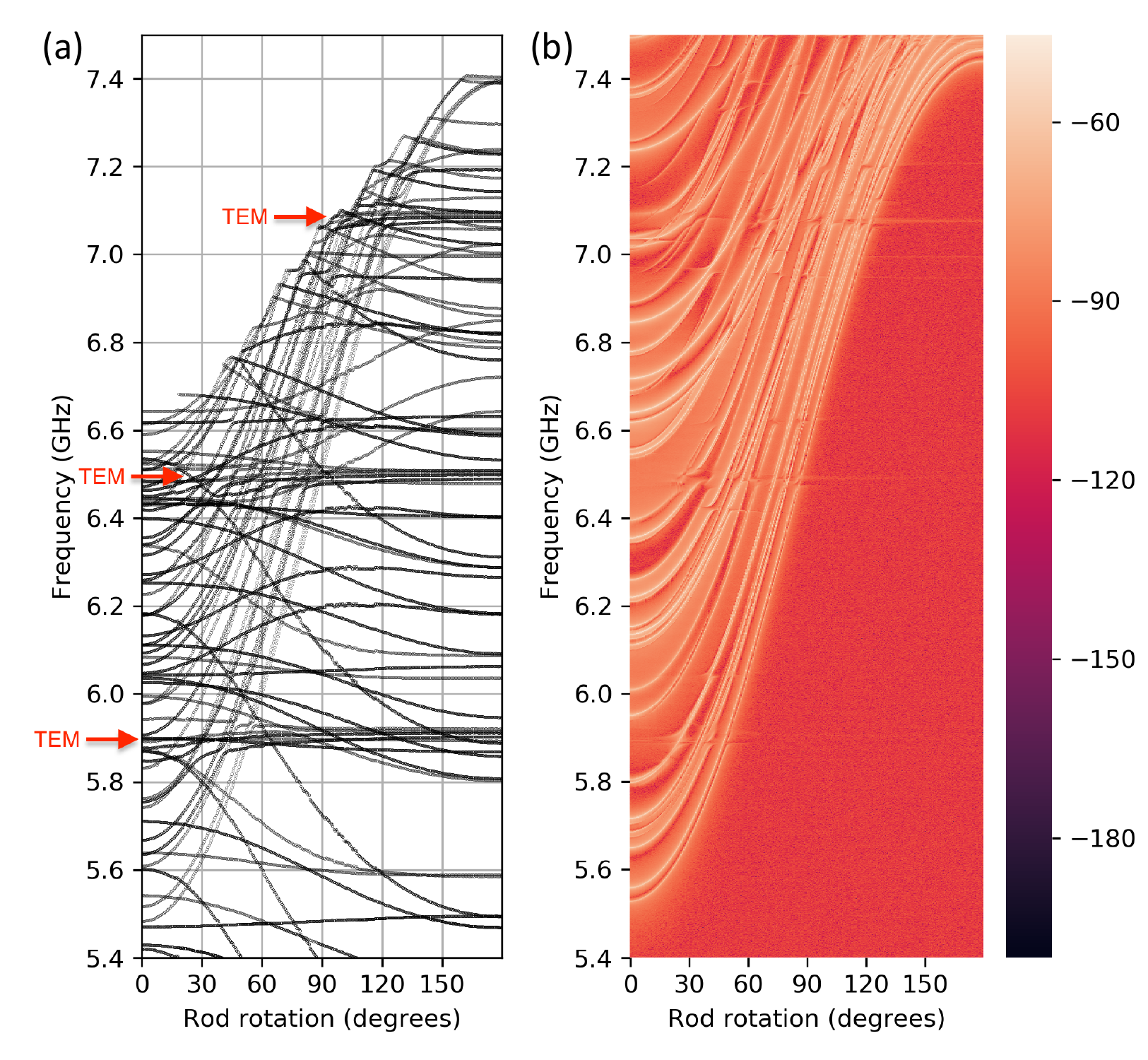}
	\caption{\label{fig:7rodmodemap}Seven $0.625''$ OD rod design mode map based on (a) eigenmode simulations and (b) measured $S_{21}$ scattering parameters. TM modes are the most prominent types of modes in the measured mode map since TE and TEM modes do not couple to our antennas as well as the TM modes. Detailed maps of a few mode crossings are presented in Fig.~\ref{fig:xing_zooms}.}
\end{figure}

\subsection{Tuning range}
The \TMone-like resonant mode frequency is tunable over $5.529 - \SI{7.440}{\giga \Hz}$ in the copper-plated seven-rod cavity. The simulated and measured mode maps in Fig.~\ref{fig:7rodmodemap} show several resonant mode frequencies as a function of rod rotation away from the innermost position. The measured mode map gives a slightly higher frequency tuning range for the \TMone-like mode than the simulated case since the copper coating was thicker than designed. Heat maps of the electric field at several of these rotation locations are shown in Fig.~\ref{fig:7rodrotations}. The measured mode map is generated by recording the $S_{21}$ scattering parameter at each half-degree rotation step of the rods and plotting the signal magnitude as a heat map. Lighter areas correspond to higher magnitude of signals and therefore resonant mode frequencies. These measurements are only sensitive to resonant modes that couple to our antennas, so we are not able to see the TE or TEM modes until they couple to the TM modes and cause mode crossings. In Fig.~\ref{fig:7rodmodemap}, the lowest-frequency line increasing in frequency with increasing rod angle corresponds to the \TMone-like mode.	
	
Throughout the tuning range, the \TMone-like mode crosses several intruder modes, as is apparent in the mode maps. TEM resonant mode frequencies in our cylindrical cavities with central conductors are determined only by the cavity height and therefore do not change with rod rotation. These TEM modes in the current seven-rod cavity, based on designed dimensions, are at $\SI{5.91}{\giga \Hz}$, $\SI{6.50}{\giga \Hz}$, and $\SI{7.09}{\giga \Hz}$, which correspond to the frequencies of the widest mode crossings in the seven-rod cavity. The other frequencies correspond to TE modes. Those that decrease in frequency with increasing rod angle do not appear in the single-rod cavity mode map since they are mostly confined to the space between the central rod and the outer rods. There are also TM modes that increase slowly with increasing rod rotation compared to the other TM modes. Simulations and bead perturbation measurements show that these are TM$_{31l}$ modes. 

\begin{figure}
	\includegraphics{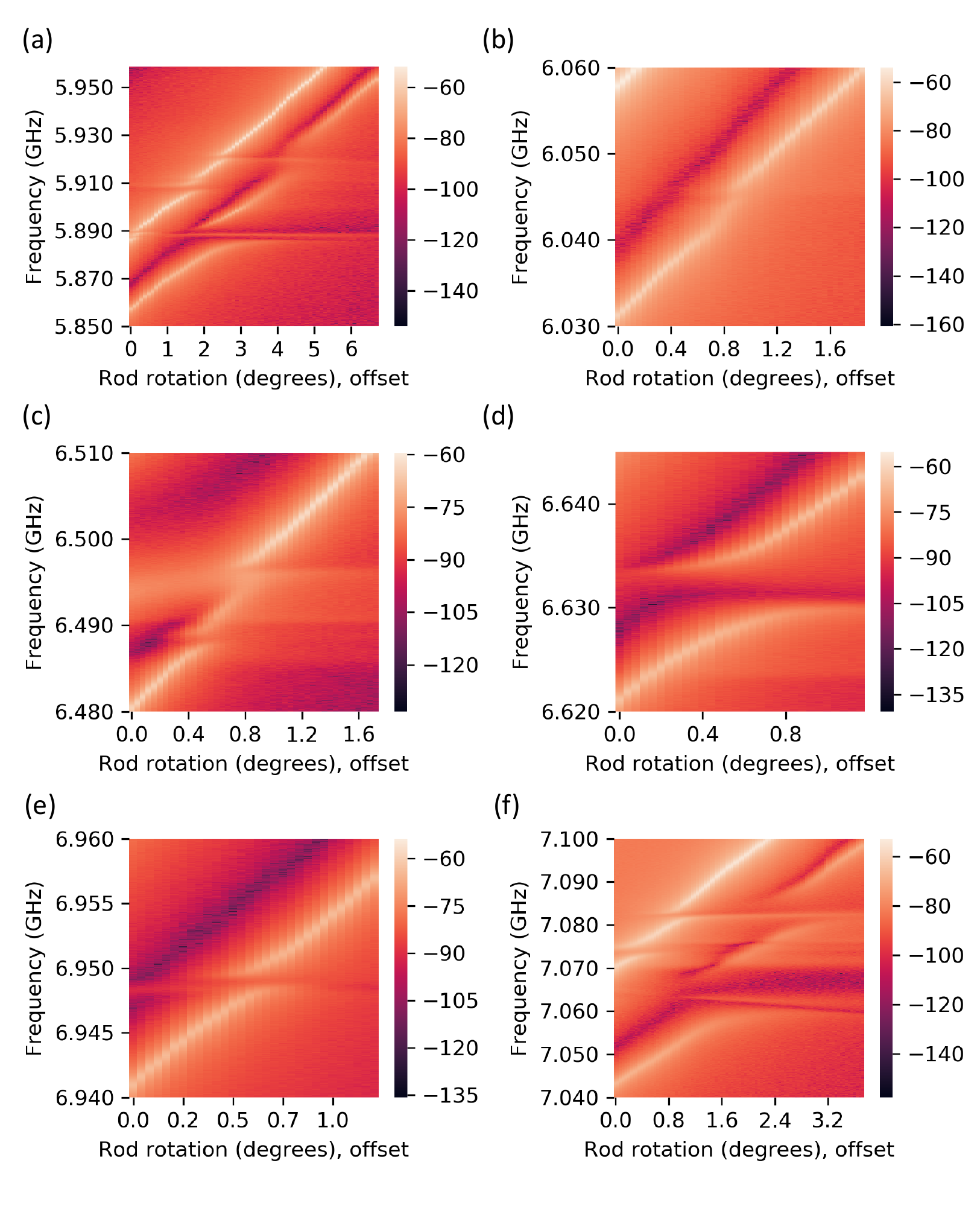}
	\caption{\label{fig:xing_zooms} Detailed mode maps of mode crossings at (a) $\SI{5.90}{\giga \Hz}$, (b) $\SI{6.04}{\giga \Hz}$, (c) $\SI{6.49}{\giga \Hz}$, (d) $\SI{6.63}{\giga \Hz}$, (e) $\SI{6.95}{\giga \Hz}$, and (f) $\SI{7.07}{\giga \Hz}$. The rod rotations are presented in degrees with an arbitrary rotation offset. As before, the associated scale gives the magnitude of the $S_{21}$ scattering parameters in units of dB. Some mode crossings are wider than others.}
\end{figure}

Figure~\ref{fig:xing_zooms} shows a few select mode crossings in detail. The lowest-frequency line increasing in frequency with increasing rod angle corresponds to the \TMone-like mode, as before. Figures~\ref{fig:xing_zooms}a, \ref{fig:xing_zooms}c, and \ref{fig:xing_zooms}f show crossings of the \TMone-like mode with TEM modes. These mode crossings occur over frequency ranges of approximately $\SI{60}{\mega \Hz}$, $\SI{20}{\mega \Hz}$, and $\SI{40}{\mega \Hz}$ at $\SI{5.90}{\giga \Hz}$, $\SI{6.49}{\giga \Hz}$, and $\SI{7.07}{\giga \Hz}$, respectively. Figures~\ref{fig:xing_zooms}b, \ref{fig:xing_zooms}d, and \ref{fig:xing_zooms}e show narrower mode crossings of approximately $\SI{5}{\mega \Hz}$, $\SI{10}{\mega \Hz}$, and $\SI{5}{\mega \Hz}$, respectively. These correspond to mode crossing of the \TMone-like mode with TE modes. There are a few other narrow mode crossings throughout the tuning range. Conservatively, we can estimate that mode crossings affect approximately $\SI{170}{\mega \Hz}$ of the $\SI{1.9}{\giga \Hz}$ tuning range of the \TMone-like mode, which means that more than $90\%$ of the tuning range $5.529 - \SI{7.440}{\giga \Hz}$ is clear of mode crossings. In the one-rod cavity, HAYSTAC used a dielectric vernier to access the frequency ranges obstructed by mode crossings, and a similar technique can be implemented in the seven-rod cavity design.

\begin{figure}
	\includegraphics[width=\textwidth]{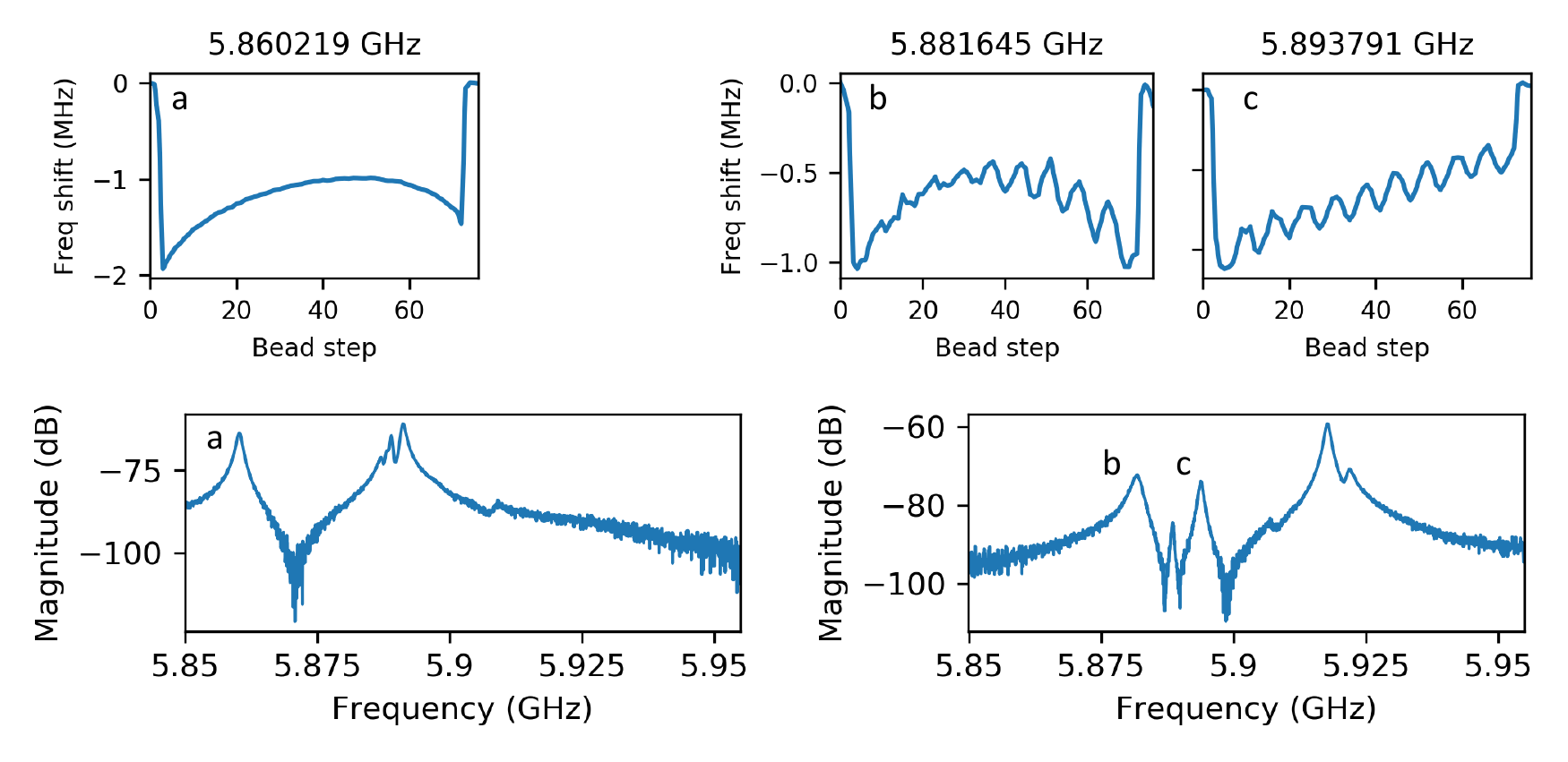}
	\caption{\label{fig:BP} Top plots show the frequency shifts observed in bead perturbation measurements to map the electric field profile of the labeled resonant modes. Bottom plots show the $S_{21}$ scattering parameter magnitude in units of dB at two rod rotations. The left-side plots give an example without mode mixing, while the right-side plots show the \TMone-like mode mixing with the tenth TEM mode at around $\SI{5.9}{\giga \Hz}$. This mode crossing also appears in Fig.~\ref{fig:xing_zooms}a.}
\end{figure}

Figure~\ref{fig:modemixing} shows a cartoon of what happens to the \TMone-like mode of interest when it approaches an intruder mode. Using a bead perturbation measurement, we can probe this interaction. Figure~\ref{fig:BP} shows the perturbation of the \TMone-like resonant mode frequency by a cylindrical dielectric alumina bead that is pulled through the length of the cavity in the third location available for antenna ports as represented by a black dot in Fig.~\ref{fig:7rodrotations}. The left-side plots in Fig.~\ref{fig:BP} include the $S_{21}$ scattering parameter showing the \TMone-like mode labeled `a' at $\SI{5.860219}{\giga \Hz}$ and its perturbed frequency throughout the bead perturbation measurement. In an ideal cavity, the \TMone-like mode electric field would not change throughout the length of the cavity, but realistically, there is some mode localization at the endcaps most likely caused by the gaps between the rods and the endcaps. The localization does not have to be symmetric, and it is not in the present case. This asymmetry is seen by the difference in the frequency shift due to the bead perturbation in the top left plot labeled `a' in Fig.~\ref{fig:BP} at bead steps of 5 and 75. In an earlier study, it was demonstrated that mode localization did not significantly impact the form factor $C$~\cite{Rap19}. The right-side plots also include the $S_{21}$ scattering parameter in the same frequency range, but at a different rod rotation. This time, the \TMone-like mode labeled `b' is at $\SI{5.881645}{\giga \Hz}$ and is mixed with the TEM mode labeled `c' at $\SI{5.893791}{\giga \Hz}$. These bead perturbation measurements show that the pure \TMone-like mode electric field shape is partially disrupted by the proximity of a TEM mode. These measurements help us confirm the mode we are investigating and determine the width of the notch in frequency which should be excised from the data analysis, as there is no longer a cleanly identified \TMone-like mode within that range. This notch is later filled in by tuning the resonant modes with the insertion of a dielectric rod, described above. In the future, we hope to be able to assess the \TMone component in a hybridized mode, benchmarking against simulations, so data of some value can be recovered from this notch.

\section{Conclusions}

HAYSTAC searches for axions between 4 and $\SI{12}{\giga \Hz}$. The \TMone-like mode in the one-rod cavity used in Phases I and II is tunable from 3.4 to $\SI{5.8}{\giga \Hz}$. The main considerations for comparing designs is the cavity figure of merit $QC^2V^2$, which depends on the quality factor, mode form factor, cavity volume, and the severity of mode crossings throughout the tuning range of the mode of interest. 

The \TMone-like mode in the two-rod designs did not reach higher frequencies effectively. The \TMone-like mode in the seven-rod cavity design is optimal for accessing 5.5 to $\SI{7.4}{\giga \Hz}$. Based on the simulation results, the seven-rod cavity was designed, built, and tested. The \TMone-like mode tuning range is over $90\%$ clear of mode crossings. 

At room temperature, the seven-rod cavity tunes smoothly and exhibits the best figure of merit of all examined designs, but is yet to be tested at cryogenic temperatures. Since HAYSTAC must reduce thermal noise contributions, the cavity must be cooled to below $\SI{100}{\milli \kelvin}$. Cooling will improve the quality factor of the \TMone-like mode in the cavity by approximately a factor of four. The effect of cooling on the mechanical motion must be tested, as well as the thermal linkage. The central rod is in contact with the endcap, and therefore is expected to cool efficiently. The other six rods have axles made of stainless steel and alumina tubes. A detailed study of inserting $0.0625''$ OD copper rods into each side of the six rod axles must be performed to test for changes in quality factor. A similar thermal link for the one-rod cavity has been well studied and successfully implemented. 

Although the seven-rod cavity provides a relatively clear tuning range of the \TMone-like mode, there are a few problematic mode crossings. In the one-rod cavity, we finely tune the \TMone-like mode resonance frequency by partially inserting a dielectric rod called the vernier. The vernier changes the frequencies of all modes and therefore the location of mode crossings. To gain access to frequencies affected by mode crossings, we must consider the possibility of incorporating a vernier into the seven-rod cavity. 

The \TMone-like mode in the current design of the seven-rod cavity reaches a frequency of around $\SI{7.4}{\giga \Hz}$, and further study will be necessary regarding the extendability of this scheme upward in frequency. Future design ideas include a seven-rod design with a differently-sized central rod than the peripheral rods. The success of the seven-rod cavity presented here gives confidence for such future similar design ideas.

\section{Acknowledgements}

This work was supported under the auspices of the National Science Foundation, under grant 1914199, and the Heising-Simons Foundation under grant 2016-044. M.S. gratefully acknowledges the University of California Berkeley Fellowship and the National Science Foundation Graduate Research Fellowship Program under grant numbers 1106400 and 1752814.

\nocite{*}
\bibliography{RSI_7rod}

\providecommand{\noopsort}[1]{}\providecommand{\singleletter}[1]{#1}%
\begin{thebibliography}{27}%
\makeatletter
\providecommand \@ifxundefined [1]{%
 \@ifx{#1\undefined}
}%
\providecommand \@ifnum [1]{%
 \ifnum #1\expandafter \@firstoftwo
 \else \expandafter \@secondoftwo
 \fi
}%
\providecommand \@ifx [1]{%
 \ifx #1\expandafter \@firstoftwo
 \else \expandafter \@secondoftwo
 \fi
}%
\providecommand \natexlab [1]{#1}%
\providecommand \enquote  [1]{``#1''}%
\providecommand \bibnamefont  [1]{#1}%
\providecommand \bibfnamefont [1]{#1}%
\providecommand \citenamefont [1]{#1}%
\providecommand \href@noop [0]{\@secondoftwo}%
\providecommand \href [0]{\begingroup \@sanitize@url \@href}%
\providecommand \@href[1]{\@@startlink{#1}\@@href}%
\providecommand \@@href[1]{\endgroup#1\@@endlink}%
\providecommand \@sanitize@url [0]{\catcode `\\12\catcode `\$12\catcode
  `\&12\catcode `\#12\catcode `\^12\catcode `\_12\catcode `\%12\relax}%
\providecommand \@@startlink[1]{}%
\providecommand \@@endlink[0]{}%
\providecommand \url  [0]{\begingroup\@sanitize@url \@url }%
\providecommand \@url [1]{\endgroup\@href {#1}{\urlprefix }}%
\providecommand \urlprefix  [0]{URL }%
\providecommand \Eprint [0]{\href }%
\providecommand \doibase [0]{http://dx.doi.org/}%
\providecommand \selectlanguage [0]{\@gobble}%
\providecommand \bibinfo  [0]{\@secondoftwo}%
\providecommand \bibfield  [0]{\@secondoftwo}%
\providecommand \translation [1]{[#1]}%
\providecommand \BibitemOpen [0]{}%
\providecommand \bibitemStop [0]{}%
\providecommand \bibitemNoStop [0]{.\EOS\space}%
\providecommand \EOS [0]{\spacefactor3000\relax}%
\providecommand \BibitemShut  [1]{\csname bibitem#1\endcsname}%
\let\auto@bib@innerbib\@empty
\bibitem [{\citenamefont {Peccei}\ and\ \citenamefont
  {Quinn}(1977{\natexlab{a}})}]{Pec77a}%
  \BibitemOpen
  \bibfield  {author} {\bibinfo {author} {\bibfnamefont {R.~D.}\ \bibnamefont
  {Peccei}}\ and\ \bibinfo {author} {\bibfnamefont {H.~R.}\ \bibnamefont
  {Quinn}},\ }\bibfield  {title} {\enquote {\bibinfo {title} {$\mathrm{CP}$
  conservation in the presence of pseudoparticles},}\ }\href {\doibase
  10.1103/PhysRevLett.38.1440} {\bibfield  {journal} {\bibinfo  {journal}
  {Phys. Rev. Lett.}\ }\textbf {\bibinfo {volume} {38}},\ \bibinfo {pages}
  {1440--1443} (\bibinfo {year} {1977}{\natexlab{a}})}\BibitemShut {NoStop}%
\bibitem [{\citenamefont {Peccei}\ and\ \citenamefont
  {Quinn}(1977{\natexlab{b}})}]{Pec77b}%
  \BibitemOpen
  \bibfield  {author} {\bibinfo {author} {\bibfnamefont {R.~D.}\ \bibnamefont
  {Peccei}}\ and\ \bibinfo {author} {\bibfnamefont {H.~R.}\ \bibnamefont
  {Quinn}},\ }\bibfield  {title} {\enquote {\bibinfo {title} {Constraints
  imposed by $\mathrm{CP}$ conservation in the presence of pseudoparticles},}\
  }\href {\doibase 10.1103/PhysRevD.16.1791} {\bibfield  {journal} {\bibinfo
  {journal} {Phys. Rev. D}\ }\textbf {\bibinfo {volume} {16}},\ \bibinfo
  {pages} {1791--1797} (\bibinfo {year} {1977}{\natexlab{b}})}\BibitemShut
  {NoStop}%
\bibitem [{\citenamefont {Weinberg}(1978)}]{Wei78}%
  \BibitemOpen
  \bibfield  {author} {\bibinfo {author} {\bibfnamefont {S.}~\bibnamefont
  {Weinberg}},\ }\bibfield  {title} {\enquote {\bibinfo {title} {A new light
  boson?}}\ }\href {\doibase 10.1103/PhysRevLett.40.223} {\bibfield  {journal}
  {\bibinfo  {journal} {Phys. Rev. Lett.}\ }\textbf {\bibinfo {volume} {40}},\
  \bibinfo {pages} {223--226} (\bibinfo {year} {1978})}\BibitemShut {NoStop}%
\bibitem [{\citenamefont {Wilczek}(1978)}]{Wil78}%
  \BibitemOpen
  \bibfield  {author} {\bibinfo {author} {\bibfnamefont {F.}~\bibnamefont
  {Wilczek}},\ }\bibfield  {title} {\enquote {\bibinfo {title} {Problem of
  strong $p$ and $t$ invariance in the presence of instantons},}\ }\href
  {\doibase 10.1103/PhysRevLett.40.279} {\bibfield  {journal} {\bibinfo
  {journal} {Phys. Rev. Lett.}\ }\textbf {\bibinfo {volume} {40}},\ \bibinfo
  {pages} {279--282} (\bibinfo {year} {1978})}\BibitemShut {NoStop}%
\bibitem [{\citenamefont {Ipser}\ and\ \citenamefont {Sikivie}(1983)}]{Sik83b}%
  \BibitemOpen
  \bibfield  {author} {\bibinfo {author} {\bibfnamefont {J.}~\bibnamefont
  {Ipser}}\ and\ \bibinfo {author} {\bibfnamefont {P.}~\bibnamefont
  {Sikivie}},\ }\bibfield  {title} {\enquote {\bibinfo {title} {Can galactic
  halos be made of axions?}}\ }\href {\doibase 10.1103/PhysRevLett.50.925}
  {\bibfield  {journal} {\bibinfo  {journal} {Phys. Rev. Lett.}\ }\textbf
  {\bibinfo {volume} {50}},\ \bibinfo {pages} {925--927} (\bibinfo {year}
  {1983})}\BibitemShut {NoStop}%
\bibitem [{\citenamefont {Stecker}\ and\ \citenamefont {Shafi}(1983)}]{Ste83}%
  \BibitemOpen
  \bibfield  {author} {\bibinfo {author} {\bibfnamefont {F.~W.}\ \bibnamefont
  {Stecker}}\ and\ \bibinfo {author} {\bibfnamefont {Q.}~\bibnamefont
  {Shafi}},\ }\bibfield  {title} {\enquote {\bibinfo {title} {Axions and the
  evolution of structure in the universe},}\ }\href {\doibase
  10.1103/PhysRevLett.50.928} {\bibfield  {journal} {\bibinfo  {journal} {Phys.
  Rev. Lett.}\ }\textbf {\bibinfo {volume} {50}},\ \bibinfo {pages} {928--931}
  (\bibinfo {year} {1983})}\BibitemShut {NoStop}%
\bibitem [{\citenamefont {{Sikivie}}(1983)}]{Sik83}%
  \BibitemOpen
  \bibfield  {author} {\bibinfo {author} {\bibfnamefont {P.}~\bibnamefont
  {{Sikivie}}},\ }\bibfield  {title} {\enquote {\bibinfo {title} {Experimental
  {Tests} of the ``{Invisible}'' {Axion}},}\ }\href {\doibase
  10.1103/PhysRevLett.51.1415} {\bibfield  {journal} {\bibinfo  {journal}
  {Phys. Rev. Lett.}\ }\textbf {\bibinfo {volume} {51}},\ \bibinfo {pages}
  {1415} (\bibinfo {year} {1983})}\BibitemShut {NoStop}%
\bibitem [{\citenamefont {Graham}\ \emph {et~al.}(2015)\citenamefont {Graham},
  \citenamefont {Irastorza}, \citenamefont {Lamoreaux}, \citenamefont
  {Lindner},\ and\ \citenamefont {van Bibber}}]{Gra15}%
  \BibitemOpen
  \bibfield  {author} {\bibinfo {author} {\bibfnamefont {P.~W.}\ \bibnamefont
  {Graham}}, \bibinfo {author} {\bibfnamefont {I.~G.}\ \bibnamefont
  {Irastorza}}, \bibinfo {author} {\bibfnamefont {S.~K.}\ \bibnamefont
  {Lamoreaux}}, \bibinfo {author} {\bibfnamefont {A.}~\bibnamefont {Lindner}},
  \ and\ \bibinfo {author} {\bibfnamefont {K.~A.}\ \bibnamefont {van Bibber}},\
  }\bibfield  {title} {\enquote {\bibinfo {title} {Experimental searches for
  the axion and axion-like particles},}\ }\href {\doibase
  10.1146/annurev-nucl-102014-022120} {\bibfield  {journal} {\bibinfo
  {journal} {Annual Review of Nuclear and Particle Science}\ }\textbf {\bibinfo
  {volume} {65}},\ \bibinfo {pages} {485--514} (\bibinfo {year} {2015})},\
  \Eprint
  {http://arxiv.org/abs/https://doi.org/10.1146/annurev-nucl-102014-022120}
  {https://doi.org/10.1146/annurev-nucl-102014-022120} \BibitemShut {NoStop}%
\bibitem [{\citenamefont {Brubaker}\ \emph
  {et~al.}(2017{\natexlab{a}})\citenamefont {Brubaker}, \citenamefont {Zhong},
  \citenamefont {Gurevich}, \citenamefont {Cahn}, \citenamefont {Lamoreaux},
  \citenamefont {Simanovskaia}, \citenamefont {Root}, \citenamefont {Lewis},
  \citenamefont {Al~Kenany}, \citenamefont {Backes}, \citenamefont {Urdinaran},
  \citenamefont {Rapidis}, \citenamefont {Shokair}, \citenamefont {van Bibber},
  \citenamefont {Palken}, \citenamefont {Malnou}, \citenamefont {Kindel},
  \citenamefont {Anil}, \citenamefont {Lehnert},\ and\ \citenamefont
  {Carosi}}]{Bru16}%
  \BibitemOpen
  \bibfield  {author} {\bibinfo {author} {\bibfnamefont {B.~M.}\ \bibnamefont
  {Brubaker}}, \bibinfo {author} {\bibfnamefont {L.}~\bibnamefont {Zhong}},
  \bibinfo {author} {\bibfnamefont {Y.~V.}\ \bibnamefont {Gurevich}}, \bibinfo
  {author} {\bibfnamefont {S.~B.}\ \bibnamefont {Cahn}}, \bibinfo {author}
  {\bibfnamefont {S.~K.}\ \bibnamefont {Lamoreaux}}, \bibinfo {author}
  {\bibfnamefont {M.}~\bibnamefont {Simanovskaia}}, \bibinfo {author}
  {\bibfnamefont {J.~R.}\ \bibnamefont {Root}}, \bibinfo {author}
  {\bibfnamefont {S.~M.}\ \bibnamefont {Lewis}}, \bibinfo {author}
  {\bibfnamefont {S.}~\bibnamefont {Al~Kenany}}, \bibinfo {author}
  {\bibfnamefont {K.~M.}\ \bibnamefont {Backes}}, \bibinfo {author}
  {\bibfnamefont {I.}~\bibnamefont {Urdinaran}}, \bibinfo {author}
  {\bibfnamefont {N.~M.}\ \bibnamefont {Rapidis}}, \bibinfo {author}
  {\bibfnamefont {T.~M.}\ \bibnamefont {Shokair}}, \bibinfo {author}
  {\bibfnamefont {K.~A.}\ \bibnamefont {van Bibber}}, \bibinfo {author}
  {\bibfnamefont {D.~A.}\ \bibnamefont {Palken}}, \bibinfo {author}
  {\bibfnamefont {M.}~\bibnamefont {Malnou}}, \bibinfo {author} {\bibfnamefont
  {W.~F.}\ \bibnamefont {Kindel}}, \bibinfo {author} {\bibfnamefont {M.~A.}\
  \bibnamefont {Anil}}, \bibinfo {author} {\bibfnamefont {K.~W.}\ \bibnamefont
  {Lehnert}}, \ and\ \bibinfo {author} {\bibfnamefont {G.}~\bibnamefont
  {Carosi}},\ }\bibfield  {title} {\enquote {\bibinfo {title} {First results
  from a microwave cavity axion search at $24\text{
  }\ensuremath{\mu}\mathrm{eV}$},}\ }\href {\doibase
  10.1103/PhysRevLett.118.061302} {\bibfield  {journal} {\bibinfo  {journal}
  {Phys. Rev. Lett.}\ }\textbf {\bibinfo {volume} {118}},\ \bibinfo {pages}
  {061302} (\bibinfo {year} {2017}{\natexlab{a}})}\BibitemShut {NoStop}%
\bibitem [{\citenamefont {Brubaker}\ \emph
  {et~al.}(2017{\natexlab{b}})\citenamefont {Brubaker}, \citenamefont {Zhong},
  \citenamefont {Lamoreaux}, \citenamefont {Lehnert},\ and\ \citenamefont {van
  Bibber}}]{Bru17}%
  \BibitemOpen
  \bibfield  {author} {\bibinfo {author} {\bibfnamefont {B.~M.}\ \bibnamefont
  {Brubaker}}, \bibinfo {author} {\bibfnamefont {L.}~\bibnamefont {Zhong}},
  \bibinfo {author} {\bibfnamefont {S.~K.}\ \bibnamefont {Lamoreaux}}, \bibinfo
  {author} {\bibfnamefont {K.~W.}\ \bibnamefont {Lehnert}}, \ and\ \bibinfo
  {author} {\bibfnamefont {K.~A.}\ \bibnamefont {van Bibber}},\ }\bibfield
  {title} {\enquote {\bibinfo {title} {Haystac axion search analysis
  procedure},}\ }\href {\doibase 10.1103/PhysRevD.96.123008} {\bibfield
  {journal} {\bibinfo  {journal} {Phys. Rev. D}\ }\textbf {\bibinfo {volume}
  {96}},\ \bibinfo {pages} {123008} (\bibinfo {year}
  {2017}{\natexlab{b}})}\BibitemShut {NoStop}%
\bibitem [{\citenamefont {Kenany}\ \emph {et~al.}(2017)\citenamefont {Kenany},
  \citenamefont {Anil}, \citenamefont {Backes}, \citenamefont {Brubaker},
  \citenamefont {Cahn}, \citenamefont {Carosi}, \citenamefont {Gurevich},
  \citenamefont {Kindel}, \citenamefont {Lamoreaux}, \citenamefont {Lehnert},
  \citenamefont {Lewis}, \citenamefont {Malnou}, \citenamefont {Palken},
  \citenamefont {Rapidis}, \citenamefont {Root}, \citenamefont {Simanovskaia},
  \citenamefont {Shokair}, \citenamefont {Urdinaran}, \citenamefont {van
  Bibber},\ and\ \citenamefont {Zhong}}]{AlK17}%
  \BibitemOpen
  \bibfield  {author} {\bibinfo {author} {\bibfnamefont {S.~A.}\ \bibnamefont
  {Kenany}}, \bibinfo {author} {\bibfnamefont {M.}~\bibnamefont {Anil}},
  \bibinfo {author} {\bibfnamefont {K.}~\bibnamefont {Backes}}, \bibinfo
  {author} {\bibfnamefont {B.}~\bibnamefont {Brubaker}}, \bibinfo {author}
  {\bibfnamefont {S.}~\bibnamefont {Cahn}}, \bibinfo {author} {\bibfnamefont
  {G.}~\bibnamefont {Carosi}}, \bibinfo {author} {\bibfnamefont
  {Y.}~\bibnamefont {Gurevich}}, \bibinfo {author} {\bibfnamefont
  {W.}~\bibnamefont {Kindel}}, \bibinfo {author} {\bibfnamefont
  {S.}~\bibnamefont {Lamoreaux}}, \bibinfo {author} {\bibfnamefont
  {K.}~\bibnamefont {Lehnert}}, \bibinfo {author} {\bibfnamefont
  {S.}~\bibnamefont {Lewis}}, \bibinfo {author} {\bibfnamefont
  {M.}~\bibnamefont {Malnou}}, \bibinfo {author} {\bibfnamefont
  {D.}~\bibnamefont {Palken}}, \bibinfo {author} {\bibfnamefont
  {N.}~\bibnamefont {Rapidis}}, \bibinfo {author} {\bibfnamefont
  {J.}~\bibnamefont {Root}}, \bibinfo {author} {\bibfnamefont {M.}~\bibnamefont
  {Simanovskaia}}, \bibinfo {author} {\bibfnamefont {T.}~\bibnamefont
  {Shokair}}, \bibinfo {author} {\bibfnamefont {I.}~\bibnamefont {Urdinaran}},
  \bibinfo {author} {\bibfnamefont {K.}~\bibnamefont {van Bibber}}, \ and\
  \bibinfo {author} {\bibfnamefont {L.}~\bibnamefont {Zhong}},\ }\bibfield
  {title} {\enquote {\bibinfo {title} {Design and operational experience of a
  microwave cavity axion detector for the 20-100 $\mu$ev range},}\ }\href
  {\doibase https://doi.org/10.1016/j.nima.2017.02.012} {\bibfield  {journal}
  {\bibinfo  {journal} {Nuclear Instruments and Methods in Physics Research
  Section A: Accelerators, Spectrometers, Detectors and Associated Equipment}\
  }\textbf {\bibinfo {volume} {854}},\ \bibinfo {pages} {11 -- 24} (\bibinfo
  {year} {2017})}\BibitemShut {NoStop}%
\bibitem [{\citenamefont {Zhong}\ \emph {et~al.}(2018)\citenamefont {Zhong},
  \citenamefont {Al~Kenany}, \citenamefont {Backes}, \citenamefont {Brubaker},
  \citenamefont {Cahn}, \citenamefont {Carosi}, \citenamefont {Gurevich},
  \citenamefont {Kindel}, \citenamefont {Lamoreaux}, \citenamefont {Lehnert},
  \citenamefont {Lewis}, \citenamefont {Malnou}, \citenamefont {Maruyama},
  \citenamefont {Palken}, \citenamefont {Rapidis}, \citenamefont {Root},
  \citenamefont {Simanovskaia}, \citenamefont {Shokair}, \citenamefont
  {Speller}, \citenamefont {Urdinaran},\ and\ \citenamefont {van
  Bibber}}]{Zho18}%
  \BibitemOpen
  \bibfield  {author} {\bibinfo {author} {\bibfnamefont {L.}~\bibnamefont
  {Zhong}}, \bibinfo {author} {\bibfnamefont {S.}~\bibnamefont {Al~Kenany}},
  \bibinfo {author} {\bibfnamefont {K.~M.}\ \bibnamefont {Backes}}, \bibinfo
  {author} {\bibfnamefont {B.~M.}\ \bibnamefont {Brubaker}}, \bibinfo {author}
  {\bibfnamefont {S.~B.}\ \bibnamefont {Cahn}}, \bibinfo {author}
  {\bibfnamefont {G.}~\bibnamefont {Carosi}}, \bibinfo {author} {\bibfnamefont
  {Y.~V.}\ \bibnamefont {Gurevich}}, \bibinfo {author} {\bibfnamefont {W.~F.}\
  \bibnamefont {Kindel}}, \bibinfo {author} {\bibfnamefont {S.~K.}\
  \bibnamefont {Lamoreaux}}, \bibinfo {author} {\bibfnamefont {K.~W.}\
  \bibnamefont {Lehnert}}, \bibinfo {author} {\bibfnamefont {S.~M.}\
  \bibnamefont {Lewis}}, \bibinfo {author} {\bibfnamefont {M.}~\bibnamefont
  {Malnou}}, \bibinfo {author} {\bibfnamefont {R.~H.}\ \bibnamefont
  {Maruyama}}, \bibinfo {author} {\bibfnamefont {D.~A.}\ \bibnamefont
  {Palken}}, \bibinfo {author} {\bibfnamefont {N.~M.}\ \bibnamefont {Rapidis}},
  \bibinfo {author} {\bibfnamefont {J.~R.}\ \bibnamefont {Root}}, \bibinfo
  {author} {\bibfnamefont {M.}~\bibnamefont {Simanovskaia}}, \bibinfo {author}
  {\bibfnamefont {T.~M.}\ \bibnamefont {Shokair}}, \bibinfo {author}
  {\bibfnamefont {D.~H.}\ \bibnamefont {Speller}}, \bibinfo {author}
  {\bibfnamefont {I.}~\bibnamefont {Urdinaran}}, \ and\ \bibinfo {author}
  {\bibfnamefont {K.~A.}\ \bibnamefont {van Bibber}},\ }\bibfield  {title}
  {\enquote {\bibinfo {title} {Results from phase 1 of the haystac microwave
  cavity axion experiment},}\ }\href {\doibase 10.1103/PhysRevD.97.092001}
  {\bibfield  {journal} {\bibinfo  {journal} {Phys. Rev. D}\ }\textbf {\bibinfo
  {volume} {97}},\ \bibinfo {pages} {092001} (\bibinfo {year}
  {2018})}\BibitemShut {NoStop}%
\bibitem [{\citenamefont {Klaer}\ and\ \citenamefont
  {Moore}(2017)}]{KlaerMoore2017}%
  \BibitemOpen
  \bibfield  {author} {\bibinfo {author} {\bibfnamefont {V.~B.}\ \bibnamefont
  {Klaer}}\ and\ \bibinfo {author} {\bibfnamefont {G.~D.}\ \bibnamefont
  {Moore}},\ }\bibfield  {title} {\enquote {\bibinfo {title} {The dark-matter
  axion mass},}\ }\href {http://stacks.iop.org/1475-7516/2017/i=11/a=049}
  {\bibfield  {journal} {\bibinfo  {journal} {Journal of Cosmology and
  Astroparticle Physics}\ }\textbf {\bibinfo {volume} {2017}},\ \bibinfo
  {pages} {049} (\bibinfo {year} {2017})}\BibitemShut {NoStop}%
\bibitem [{\citenamefont {Buschmann}, \citenamefont {Foster},\ and\
  \citenamefont {Safdi}(2020)}]{Bus20}%
  \BibitemOpen
  \bibfield  {author} {\bibinfo {author} {\bibfnamefont {M.}~\bibnamefont
  {Buschmann}}, \bibinfo {author} {\bibfnamefont {J.~W.}\ \bibnamefont
  {Foster}}, \ and\ \bibinfo {author} {\bibfnamefont {B.~R.}\ \bibnamefont
  {Safdi}},\ }\bibfield  {title} {\enquote {\bibinfo {title} {Early-universe
  simulations of the cosmological axion},}\ }\href {\doibase
  10.1103/PhysRevLett.124.161103} {\bibfield  {journal} {\bibinfo  {journal}
  {Phys. Rev. Lett.}\ }\textbf {\bibinfo {volume} {124}},\ \bibinfo {pages}
  {161103} (\bibinfo {year} {2020})}\BibitemShut {NoStop}%
\bibitem [{\citenamefont {Peng}\ \emph {et~al.}(2000)\citenamefont {Peng} \emph
  {et~al.}}]{Pen00}%
  \BibitemOpen
  \bibfield  {author} {\bibinfo {author} {\bibfnamefont {H.}~\bibnamefont
  {Peng}} \emph {et~al.},\ }\bibfield  {title} {\enquote {\bibinfo {title}
  {{Cryogenic cavity detector for a large scale cold dark-matter axion
  search}},}\ }\href {\doibase 10.1016/S0168-9002(99)00971-7} {\bibfield
  {journal} {\bibinfo  {journal} {Nucl. Instrum. Meth.}\ }\textbf {\bibinfo
  {volume} {A444}},\ \bibinfo {pages} {569--583} (\bibinfo {year}
  {2000})}\BibitemShut {NoStop}%
\bibitem [{\citenamefont {Read}(2014)}]{Rea14}%
  \BibitemOpen
  \bibfield  {author} {\bibinfo {author} {\bibfnamefont {J.~I.}\ \bibnamefont
  {Read}},\ }\bibfield  {title} {\enquote {\bibinfo {title} {The local dark
  matter density},}\ }\href {\doibase 10.1088/0954-3899/41/6/063101} {\bibfield
   {journal} {\bibinfo  {journal} {Journal of Physics G: Nuclear and Particle
  Physics}\ }\textbf {\bibinfo {volume} {41}},\ \bibinfo {pages} {063101}
  (\bibinfo {year} {2014})}\BibitemShut {NoStop}%
\bibitem [{\citenamefont {Dicke}(1946)}]{Dic46}%
  \BibitemOpen
  \bibfield  {author} {\bibinfo {author} {\bibfnamefont {R.~H.}\ \bibnamefont
  {Dicke}},\ }\bibfield  {title} {\enquote {\bibinfo {title} {The measurement
  of thermal radiation at microwave frequencies},}\ }\href@noop {} {\bibfield
  {journal} {\bibinfo  {journal} {Review of Scientific Instruments}\ }\textbf
  {\bibinfo {volume} {17}},\ \bibinfo {pages} {268--275} (\bibinfo {year}
  {1946})}\BibitemShut {NoStop}%
\bibitem [{\citenamefont {Malnou}\ \emph {et~al.}(2019)\citenamefont {Malnou},
  \citenamefont {Palken}, \citenamefont {Brubaker}, \citenamefont {Vale},
  \citenamefont {Hilton},\ and\ \citenamefont {Lehnert}}]{Mal19}%
  \BibitemOpen
  \bibfield  {author} {\bibinfo {author} {\bibfnamefont {M.}~\bibnamefont
  {Malnou}}, \bibinfo {author} {\bibfnamefont {D.~A.}\ \bibnamefont {Palken}},
  \bibinfo {author} {\bibfnamefont {B.~M.}\ \bibnamefont {Brubaker}}, \bibinfo
  {author} {\bibfnamefont {L.~R.}\ \bibnamefont {Vale}}, \bibinfo {author}
  {\bibfnamefont {G.~C.}\ \bibnamefont {Hilton}}, \ and\ \bibinfo {author}
  {\bibfnamefont {K.~W.}\ \bibnamefont {Lehnert}},\ }\bibfield  {title}
  {\enquote {\bibinfo {title} {Squeezed vacuum used to accelerate the search
  for a weak classical signal},}\ }\href {\doibase 10.1103/PhysRevX.9.021023}
  {\bibfield  {journal} {\bibinfo  {journal} {Phys. Rev. X}\ }\textbf {\bibinfo
  {volume} {9}},\ \bibinfo {pages} {021023} (\bibinfo {year}
  {2019})}\BibitemShut {NoStop}%
\bibitem [{\citenamefont {Pippard}(1947)}]{Pip47}%
  \BibitemOpen
  \bibfield  {author} {\bibinfo {author} {\bibfnamefont {A.~B.}\ \bibnamefont
  {Pippard}},\ }\bibfield  {title} {\enquote {\bibinfo {title} {The surface of
  superconductors and normal metals at high frequencies ii. the anomalous skin
  effect in normal metals},}\ }\href {\doibase
  https://doi.org/10.1098/rspa.1947.0122} {\bibfield  {journal} {\bibinfo
  {journal} {Proc. R. Soc. Lond. A.}\ }\textbf {\bibinfo {volume} {191}},\
  \bibinfo {pages} {385--399} (\bibinfo {year} {1947})}\BibitemShut {NoStop}%
\bibitem [{\citenamefont {Kittel}(1987)}]{Kit87}%
  \BibitemOpen
  \bibfield  {author} {\bibinfo {author} {\bibfnamefont {C.}~\bibnamefont
  {Kittel}},\ }\href@noop {} {\emph {\bibinfo {title} {Quantum Theory of
  Solids}}}\ (\bibinfo  {publisher} {J. Wiley and Sons},\ \bibinfo {year}
  {1987})\BibitemShut {NoStop}%
\bibitem [{Note1()}]{Note1}%
  \BibitemOpen
  \bibinfo {note} {Although this volume definition differs from a
  previously-used definition, the scan rate calculations are equivalent since
  volume only appears as a product with the form factor. Other experiments
  defined volume as the space inside the cavity outer metal boundary, including
  the metal pieces as well as the space involving vacuum and dielectric
  materials.}\BibitemShut {Stop}%
\bibitem [{\citenamefont {Rapidis}, \citenamefont {Lewis},\ and\ \citenamefont
  {van Bibber}(2019)}]{Rap19}%
  \BibitemOpen
  \bibfield  {author} {\bibinfo {author} {\bibfnamefont {N.~M.}\ \bibnamefont
  {Rapidis}}, \bibinfo {author} {\bibfnamefont {S.~M.}\ \bibnamefont {Lewis}},
  \ and\ \bibinfo {author} {\bibfnamefont {K.~A.}\ \bibnamefont {van Bibber}},\
  }\bibfield  {title} {\enquote {\bibinfo {title} {Characterization of the
  haystac axion dark matter search cavity using microwave measurement and
  simulation techniques},}\ }\href {\doibase 10.1063/1.5055246} {\bibfield
  {journal} {\bibinfo  {journal} {Review of Scientific Instruments}\ }\textbf
  {\bibinfo {volume} {90}},\ \bibinfo {pages} {024706} (\bibinfo {year}
  {2019})}\BibitemShut {NoStop}%
\bibitem [{Note2()}]{Note2}%
  \BibitemOpen
  \bibinfo {note} {Computer Simulation Technology Microwave Studio,
  https://www.cst.com/products/cstmws}\BibitemShut {NoStop}%
\bibitem [{Note3()}]{Note3}%
  \BibitemOpen
  \bibinfo {note} {Stock Drive Products / Sterling Instrument,
  https://sdp-si.com/}\BibitemShut {NoStop}%
\bibitem [{\citenamefont {Maier}\ and\ \citenamefont {Slater}(1952)}]{Mai52}%
  \BibitemOpen
  \bibfield  {author} {\bibinfo {author} {\bibfnamefont {L.~C.}\ \bibnamefont
  {Maier}}\ and\ \bibinfo {author} {\bibfnamefont {J.~C.}\ \bibnamefont
  {Slater}},\ }\bibfield  {title} {\enquote {\bibinfo {title} {Field strength
  measurements in resonant cavities},}\ }\href {\doibase 10.1063/1.1701980}
  {\bibfield  {journal} {\bibinfo  {journal} {Journal of Applied Physics}\
  }\textbf {\bibinfo {volume} {23}},\ \bibinfo {pages} {68--77} (\bibinfo
  {year} {1952})}\BibitemShut {NoStop}%
\bibitem [{\citenamefont {Slater}(1946)}]{Sla46}%
  \BibitemOpen
  \bibfield  {author} {\bibinfo {author} {\bibfnamefont {J.~C.}\ \bibnamefont
  {Slater}},\ }\bibfield  {title} {\enquote {\bibinfo {title} {Microwave
  electronics},}\ }\href {\doibase 10.1103/RevModPhys.18.441} {\bibfield
  {journal} {\bibinfo  {journal} {Rev. Mod. Phys.}\ }\textbf {\bibinfo {volume}
  {18}},\ \bibinfo {pages} {441--512} (\bibinfo {year} {1946})}\BibitemShut
  {NoStop}%
\bibitem [{\citenamefont {Hagmann}\ \emph {et~al.}(1990)\citenamefont
  {Hagmann}, \citenamefont {Sikivie}, \citenamefont {Sullivan}, \citenamefont
  {Tanner},\ and\ \citenamefont {Cho}}]{Hag90}%
  \BibitemOpen
  \bibfield  {author} {\bibinfo {author} {\bibfnamefont {C.}~\bibnamefont
  {Hagmann}}, \bibinfo {author} {\bibfnamefont {P.}~\bibnamefont {Sikivie}},
  \bibinfo {author} {\bibfnamefont {N.}~\bibnamefont {Sullivan}}, \bibinfo
  {author} {\bibfnamefont {D.~B.}\ \bibnamefont {Tanner}}, \ and\ \bibinfo
  {author} {\bibfnamefont {S.-I.}\ \bibnamefont {Cho}},\ }\bibfield  {title}
  {\enquote {\bibinfo {title} {Cavity design for a cosmic axion detector},}\
  }\href {\doibase 10.1063/1.1141427} {\bibfield  {journal} {\bibinfo
  {journal} {Review of Scientific Instruments}\ }\textbf {\bibinfo {volume}
  {61}},\ \bibinfo {pages} {1076--1085} (\bibinfo {year} {1990})}\BibitemShut
  {NoStop}%
\end{thebibliography}%

\end{document}